 \documentclass[prd,aps,
twocolumn,
nofootinbib,
floatfix,
superscriptaddress]{revtex4}
\usepackage{graphicx}
\usepackage{epsfig}
\usepackage{rotating}
\usepackage{amssymb}
\usepackage{subfigure}
\usepackage{dsfont}
\usepackage{psfrag}
\usepackage{amsmath,euscript,array,mathrsfs}
\usepackage{bbold}
\usepackage{xcolor}

\newcommand{\comm}[1]{} 
 
\newcommand{\beq}{\begin{equation}}
\newcommand{\eeq}{\end{equation}}
\newcommand{\beqs}{\begin{eqnarray}}
\newcommand{\eeqs}{\end{eqnarray}}

\newcommand{\Tr}{{\rm Tr}}

\def\hbar{\hspace{0pt}\raisebox{1pt}{$-$} \hspace{-7pt} h}

\def\di{\mbox{d}}
\def\r{\rho}

\newcommand{\be}{\begin{equation}}
\newcommand{\ee}{\end{equation}}
\newcommand{\bea}{\begin{eqnarray}}
\newcommand{\eea}{\end{eqnarray}}

\def\co{{\cal O}}

\def\lbldef#1#2{\expandafter\gdef\csname #1\endcsname {#2}}

\def\href#1#2{#2}

\newcommand{\ber}{\begin{eqnarray}}
\newcommand{\eer}{\end{eqnarray}}

\newcommand{\beqar}{\begin{eqnarray}}

\newcommand{\eeqar}{\end{eqnarray}}

\newcommand{\dsl}{\kern.06em\hbox{\raise.15ex\hbox{$/$}\kern-.56em\hbox{$\partial$}}}

\newcommand{\eeqarr}{\end{eqnarray}}
\newcommand{\ZZ}{{\rm \kern 0.275em Z \kern -0.92em Z}\;}

\def\CC{{\mathchoice
{\rm C\mkern-8mu\vrule height1.45ex depth-.05ex
width.05em\mkern9mu\kern-.05em}
{\rm C\mkern-8mu\vrule height1.45ex depth-.05ex
width.05em\mkern9mu\kern-.05em}
{\rm C\mkern-8mu\vrule height1ex depth-.07ex
width.035em\mkern9mu\kern-.035em}
{\rm C\mkern-8mu\vrule height.65ex depth-.1ex
width.025em\mkern8mu\kern-.025em}}}

\def\RR{{\rm I\kern-1.6pt {\rm R}}}

\def\ZZ{{\rm Z}\kern-3.8pt {\rm Z} \kern2pt}
\def\IB{\relax{\rm I\kern-.18em B}}
\def\ID{\relax{\rm I\kern-.18em D}}
\def\II{\relax{\rm I\kern-.18em I}}
\def\IP{\relax{\rm I\kern-.18em P}}

\newcommand{\bear}{\begin{eqnarray}}
\newcommand{\eear}{\end{eqnarray}}

\def\r{\rho}

\def\6{\partial}

\def\bea{\begin{eqnarray}}
\def\eea{\end{eqnarray}}

\def\beqx{\begin{displaymath}}
\def\eeqx{\end{displaymath}}

\newcommand{\bmat}{\left(\begin{array}}
\newcommand{\emat}{\end{array}\right)}

\def\r{\rho}

\def\co{{\cal O}}

\def\bo{{\raise-.3ex\hbox{\large$\Box$}}}               
\def\face{{\raise.2ex\hbox{$\displaystyle \bigodot$}\mskip-2.2mu \llap {$\ddot
        \smile$}}}                                   
\def\>{\rangle}                                      
\def\<{\langle}                                      

\def\leftrightarrowfill{$\mathsurround=0pt \mathord\leftarrow \mkern-6mu
        \cleaders\hbox{$\mkern-2mu \mathord- \mkern-2mu$}\hfill
        \mkern-6mu \mathord\rightarrow$}        
\def\dvec#1{\vbox{\ialign{##\crcr
        \leftrightarrowfill\crcr\noalign{\kern-1pt\nointerlineskip}
        $\hfil\displaystyle{#1}\hfil$\crcr}}}           
\def\Tr{{\rm Tr \,}}                                    

\def\-{\hphantom{-}}

\begin{document}

\title{The Coulomb branch of ${\cal N}=4$ SYM and dilatonic scions in supergravity
}
\author{Daniel Elander}
\affiliation{Laboratoire Charles Coulomb (L2C), University of Montpellier, CNRS, Montpellier, France}

\author{Maurizio Piai}
\author{John Roughley}

\affiliation{Department of Physics, College of Science, Swansea University,
Singleton Park, SA2 8PP, Swansea, Wales, UK}

\date{\today}


\begin{abstract}

  We find a parametrically light dilaton in special confining theories in three dimensions. Their duals form what we call a \emph{scion} of solutions to the supergravity associated with the large-$N$ limit of the Coulomb branch of the ${\cal N}=4$ Super-Yang-Mills (SYM) theory. 
 The supergravity description contains one scalar with bulk mass that saturates the 
  Breitenlohner-Freedman unitarity bound.
 The new solutions are defined within supergravity, they
 break supersymmetry and scale invariance, and one
	  dimension is compactified on a shrinking circle, yet they are completely regular.
	 An approximate dilaton appears in the spectrum of background
	 fluctuations (or composite states in the confining theory), and becomes parametrically light
	 along a metastable portion of the scion of new supergravity solutions,
	in close proximity of a tachyonic instability. A first-order phase transition separates
	stable backgrounds, for which the approximate dilaton is not parametrically light,
	from metastable and unstable backgrounds, for which the dilaton becomes parametrically light,
	and eventually tachyonic.

\end{abstract}

\maketitle

\tableofcontents

\allowdisplaybreaks

\section{Introduction}
\label{Sec:Introduction}

In Refs.~\cite{ Elander:2020ial} and~\cite{Elander:2020fmv}, we proposed a  mechanism 
giving rise to an approximate dilaton in the spectrum of composite states of special classes
of confining theories, in four dimensions, that
admit a higher-dimensional gravity dual.
We provided two explicit, calculable realisations of this mechanism,
which generalises the ideas proposed in Ref.~\cite{Kaplan:2009kr} (and further
critically discussed in Refs.~\cite{Gorbenko:2018ncu,Gorbenko:2018dtm,Pomarol:2019aae}).
We considered non-AdS gravity backgrounds 
in proximity of classical instabilities---generalising the proximity to the 
 Breitenlohner-Freedman (BF) unitarity  bound~\cite{Breitenlohner:1982jf}
in AdS space.
An approximate
 dilaton (a scalar particle coupling to the trace of the energy-momentum tensor) 
 emerges along special branches of  supergravity solutions. Portions of the branches
 yield stable solutions, while complementary ones
 describe metastable  or even unstable solutions.
Moving in parameter space along the branch of solutions,  the dilaton 
has finite mass for stable solutions, becomes parametrically
 light for metastable ones, and tachyonic for the unstable ones.

The analysis follows the prescriptions of
gauge-gravity dualities~\cite{Maldacena:1997re,Gubser:1998bc,Witten:1998qj,Aharony:1999ti},
in the calculation 
of the free energy, via holographic renormalisation~\cite{Bianchi:2001kw,Skenderis:2002wp,Papadimitriou:2004ap},
and of the spectrum of the bound states, via a convenient gauge-invariant 
formalism~\cite{Bianchi:2003ug,Berg:2005pd,Berg:2006xy,Elander:2009bm,Elander:2010wd}.
We adopt the holographic description of confinement, the calculation of
Wilson loops~\cite{Rey:1998ik,Maldacena:1998im,Kinar:1998vq,
Brandhuber:1999jr,Avramis:2006nv,Nunez:2009da,Faedo:2013ota},
and the interpretation of singularities~\cite{Gubser:2000nd}.
We restrict attention  to well established supergravity theories (top-down
 holography).
Ref.~\cite{Elander:2020ial} studies the
 half-maximal supergravity in $D=6$ dimensions, due to Romans~\cite{Romans:1985tw,Romans:1985tz,
 Brandhuber:1999np,Cvetic:1999un} (see also 
 Refs.~\cite{Hong:2018amk,Jeong:2013jfc,DAuria:2000afl,
 Andrianopoli:2001rs,Nishimura:2000wj,Ferrara:1998gv,Gursoy:2002tx,Nunez:2001pt,Karndumri:2012vh,
 Lozano:2012au,Karndumri:2014lba,Chang:2017mxc,Gutperle:2018axv,Kim:2019fsg,
 Chen:2019qib,Hoyos:2020fjx,Suh:2018tul,Suh:2018szn}), compactified on a circle~\cite{Wen:2004qh,
 Kuperstein:2004yf,Elander:2013jqa,Elander:2018aub},
while Ref.~\cite{Elander:2020fmv} considers the maximal supergravity in $D=7$ 
 dimensions~\cite{Nastase:1999cb,Pernici:1984xx,Pernici:1984zw,Lu:1999bc,
 Campos:2000yu},
 compactified on a 2-torus~\cite{Witten:1998zw,Brower:2000rp}.
In Refs.~\cite{Elander:2020ial,Elander:2020fmv}, we explored regions of the admissible parameter space overlooked 
 in the earlier literature.

The significance of Refs.~\cite{ Elander:2020ial,Elander:2020fmv}
extends beyond producing calculable examples of the ideas exposed in
  Refs.~\cite{Kaplan:2009kr,Gorbenko:2018ncu,Gorbenko:2018dtm,Pomarol:2019aae}.
Recent years saw a resurgence of interest  in the literature on the dilaton EFT 
 description of near-conformal, strongly coupled systems in four  dimensions
  (see for instance Refs.~\cite{Matsuzaki:2013eva,
 Golterman:2016lsd,Kasai:2016ifi,Golterman:2016hlz,Hansen:2016fri,Golterman:2016cdd,Appelquist:2017wcg,
 Appelquist:2017vyy,Golterman:2018mfm,Cata:2019edh,Appelquist:2019lgk,Cata:2018wzl,
 Brown:2019ipr}). 
 The literature on the subject has a long history~\cite{Coleman:1985rnk}, including early attempts to 
 explain the long distance behaviour of Yang-Mills~\cite{Migdal:1982jp}
and walking technicolor~\cite{Leung:1985sn,Bardeen:1985sm,Yamawaki:1985zg} theories.
One motivation for the revival is the uncovering of
 a light scalar particle, possibly a dilaton, in special $SU(3)$ lattice gauge theories coupled to 
 matter~\cite{Aoki:2014oha,Appelquist:2016viq, 
 Aoki:2016wnc,Gasbarro:2017fmi,Appelquist:2018yqe,Fodor:2012ty,Fodor:2015vwa,
 Fodor:2016pls,Fodor:2017nlp,Fodor:2019vmw,Fodor:2020niv,Golterman:2020tdq}.
Previously, the study of
simple bottom-up holographic models showed the emergence of a dilaton in special cases
of theoretical  relevance~\cite{Goldberger:1999uk,
DeWolfe:1999cp,Goldberger:1999un,Csaki:2000zn,ArkaniHamed:2000ds,Rattazzi:2000hs,Kofman:2004tk,
Elander:2011aa,Kutasov:2012uq,Lawrance:2012cg,Elander:2012fk,Goykhman:2012az,
Evans:2013vca,Megias:2014iwa,Elander:2015asa}.
Even  top-down holography provided supporting 
 evidence for the existence of a light scalar in special confining 
 theories~\cite{ Elander:2009pk,Elander:2012yh,Elander:2014ola,
 Elander:2017cle,Elander:2017hyr}
 related to the conifold~\cite{Candelas:1989js,Chamseddine:1997nm,Klebanov:1998hh,
 Klebanov:2000hb, Maldacena:2000yy, Butti:2004pk,Nunez:2008wi}.
Furthermore, the distinctive features of the dilaton EFT
have been applied to phenomenological 
extensions of 
 the standard model~\cite{Goldberger:2008zz,Hong:2004td,Dietrich:2005jn,Hashimoto:2010nw,
 Appelquist:2010gy,Vecchi:2010gj,Chacko:2012sy,Bellazzini:2012vz,Abe:2012eu,
 Eichten:2012qb,Bellazzini:2013fga,Hernandez-Leon:2017kea},  including their use
in the context of composite Nambu-Goldstone-Higgs models~\cite{Appelquist:2020bqj}.

The gravity duals of the confining theories studied in 
Refs.~\cite{ Elander:2020ial,Elander:2020fmv} exhibit large departures
from AdS geometry. What renders one of the scalar fluctuations
 parametrically light along the metastable branch is 
the interplay between the presence of a vacuum expectation value (VEV)
 breaking spontaneously scale invariance,  the explicit breaking due to relevant deformations,
 and the effects of the nearby instability. The resulting scalar is an
 approximate dilaton, in the sense that it couples as expected to the trace of the
 energy-momentum tensor of the dual field theory; this is demonstrated by the failure of
  the probe approximation (which ignores the fluctuations of the trace of metric~\cite{Elander:2020csd})
 to reproduce correctly 
 the mass spectrum.
 We refer the reader to the original 
publications for the details, and we  defer commenting on potential
 phenomenological applications to extensions of the standard model and to
  Higgs physics~\cite{Aad:2012tfa,Chatrchyan:2012xdj}.

The purpose of this paper is to exhibit a third example of this mechanism,  
but realised in a lower dimensional theory, in backgrounds that in the far UV approach an 
AdS geometry, with bulk scalar mass close to the BF bound. This hence highlights differences and similarities
with other proposals for the origin of the dilaton.
We study a particular truncation of the ${\cal N}=8$ maximal supergravity theory in $D=5$ dimensions,
which is (loosely) associated with the Coulomb branch of
${\cal N}=4$ super-Yang-Mills (SYM) theories.
By introducing a relevant deformation, and compactifying one dimension on a circle, we build
a scion\footnote{
 The  dual of the Coulomb branch is sourced by a discrete distribution of displaced D3 branes; conversely
the pure supergravity action and lift we borrow from the literature~\cite{Pilch:2000ue} 
leads to singular supersymmetric solutions. 
 In view of this loose relation between supergravity theory and Coulomb branch, 
 we refer to our new solutions, obtained by elaborating on the gravity theory,  
 as forming a scion, rather than a branch, as a way  to emphasize their hybrid nature.}
  of gravity backgrounds yielding a light dilaton and
 admitting an interpretation in terms of a field theory  in three dimensions that 
 confines.
The scion provides a one-parameter family generalisation of the gravity background occasionally
denoted in the literature as  QCD$_3$, and for which the spectrum of fluctuations
is known~\cite{Brower:2000rp}.

The paper is organised as follows.
We  summarise in Sect.~\ref{Sec:Model} the main features of the Coulomb branch,
as well as the consistent truncation of maximal  supergravity
in $D=5$ dimensions, its reduction to $D=4$ dimensions, the lift to $D=10$ type IIB supergravity,
and the prescription for Wilson loops.
Sect.~\ref{Sec:Bg} summarises the classes of solutions we investigate in this paper: we present the UV
 and IR expansions, then compute curvature invariants and Wilson loops to characterise the solutions.
 We compute the spectrum of fluctuations for the  solutions in Sect.~\ref{Sec:spectrum}, 
 in  the appropriate number of dimensions.
In Sect.~\ref{Sec:Comparison} we compare the free energies, to discuss the  stability of the solutions.
 After the conclusions in Sect.~\ref{Sec:Conclusions}, we supplement the material with 
 Appendix~\ref{Sec:susy}, summarising known results for the supersymmetric solutions,
 and Appendix~\ref{Sec:exp}, exhibiting asymptotic expansions of the fluctuations
 used in computing the spectra.

\section{The model}
\label{Sec:Model}

 The ${\cal N}=8$ maximal supergravity in $D=5$ dimensions~\cite{Pernici:1985ju,Gunaydin:1984qu,
 Gunaydin:1985cu}
  has played a central role 
 in the history of gauge-gravity dualities.
It descends from dimensional reduction of 
 type IIB supergravity in   $D=10$ dimensions on the 5-sphere $S^5$~\cite{Gunaydin:1984fk,Kim:1985ez}. 
 It has recently been established that this is a consistent truncation~\cite{Lee:2014mla,Baguet:2015sma}, and the full uplift
back to type IIB  is known~\cite{Baguet:2015sma,Hohm:2013vpa,Baguet:2015xha}
  (see also Refs.~\cite{Cvetic:2000nc,Pilch:2000ue,Bakas:1999ax}).
The gauge symmetry is
$SO(6)\sim SU(4)$---capturing the isometries of $S^5$, and the R-symmetry of the dual field theory, respectively.
 The field content includes  42 real scalars that  match 
 ${\cal N}=4$ field-theory operators
  on the basis of their transformation properties under $SU(4)$:
the complex singlet   $\mathbb{1}_{\mathbb{C}}$
 corresponds to the holomorphic gauge coupling, 
 the symmetric $10_{\mathbb{C}}$ to the fermion masses, and the real $20^{\prime}$ 
 to the matrix of squared masses for the scalars $X_i$, with $i=1\,,\cdots\,,6$
  (see e.g.  Sect.~2.2.5 of Ref.~\cite{Aharony:1999ti},
or the introduction of  Ref.~\cite{Distler:1998gb}). 
 
  One of the  background solutions of $D=5$ maximal supergravity 
 lifts in type IIB to the AdS$_{5}\times S^5$ background geometry providing the weakly-coupled
  dual description of ${\cal N}=4$ SYM 
 with $SU(N)$ gauge group, in the (decoupling) limit of large $N$ and large 't Hooft coupling~\cite{Maldacena:1997re}.
 The supergravity solution is also the appropriate decoupling limit of 
  the configuration sourced by a stack of $N$ coincident  D3 branes.
Following Ref.~\cite{Freedman:1999gk} (see also Refs.~\cite{Kraus:1998hv,Brandhuber:1999jr,Cvetic:1999xx}), 
we call Coulomb branch the space of inequivalent
vacua of the ${\cal N}=4$ theory that preserve $16$ supercharges.
The space is so called because 
 away from its $SO(6)$-invariant configuration 
 the gauge group of the field theory is partially higgsed, and
the massless gauge bosons mediate Coulomb interactions.

In the 
language of extended objects in $D=10$ dimensions, the literature identified
multi-centred D3-brane solutions~\cite{Freedman:1999gk,Brandhuber:1999jr}
with the moduli space of ${\cal N}=4$ SYM, in the sense that  points of
the Coulomb branch  are associated with distributions
of the $N$ D3 branes over $\mathbb{R}^6$
(conveniently parametrised as a cone over the sphere $S^5$),
 accompanied by the higgsing of $SU(N)$.
By taking $N\rightarrow \infty$,  while introducing a continuous distribution 
of D3 branes, one might hope to recover a supergravity description of the 
Coulomb branch still within maximal ${\cal N}=8$ supergravity in $D=5$ dimensions. 
In fact, the resulting metrics satisfy the supergravity equations~\cite{Freedman:1999gk},
but are singular (see for example
 the discussion after Eq.~(3.12) of  Ref.~\cite{Hernandez:2005xd}).
These solutions
are captured by  a consistent truncation~\cite{Cvetic:2000nc,Pilch:2000ue,Bakas:1999ax}  that retains 
 only the $20^{\prime}$ scalars,
   dual to the symmetric and traceless operator
   \beqs
20^{\prime} \,\sim\,  \left(\delta_i^{\,\,k}\delta_j^{\,\,\ell}-\frac{1}{6}\delta_{ij}\delta^{kl}\right)\Tr X_kX_{\ell}\,.
   \eeqs
There are  five subclasses of  solutions
 that preserve $SO(n)\times SO(6-n)$ subgroups of $SO(6)$, with $n=1,\cdots,5$~\cite{Freedman:1999gk}.
With abuse of language,  the supersymmetric backgrounds of this type
are referred to as the Coulomb branch, though such solutions are
 singular, and hence incomplete as gravity duals.

We further restrict our attention  to the $n=2$ 
and the $n=4$ cases~\cite{Brandhuber:1999jr}.
The spectrum of the $n=2$ case is quite peculiar:
 both the spin-0 and spin-2 spectra have a gap and a cut opening above a finite value~\cite{
 Freedman:1999gk,
 Brandhuber:1999hb,Bianchi:2000sm,Brandhuber:2000fr,
 Bianchi:2001kw,Papadimitriou:2004rz} (see also
 Refs.~\cite{Brandhuber:2000fr,Brandhuber:2002rx} for the spectra of vectors).
 The choices $n=2,4$ are convenient also because they are both captured by
 one of the subtruncations of the theory in Ref.~\cite{Pilch:2000fu}---which retains only 
 two scalars, one in  the $20^{\prime}$ and the $10_{\mathbb{C}}$, respectively
   (see also the discussions in Refs.~\cite{Khavaev:1998fb,Freedman:1999gp}).
Setting to zero the latter of the two scalars reduces the field content to just one scalar ($\phi$ in our notation),
and the lift to $D=10$ dimensions is 
comparatively simple.

The gravity descriptions for $n=2,4$ are different;
 $n$ is associated with the ball ${\cal B}^n$
 inside the internal space (including the radial direction) over which one distributes the $N$
  D3 branes, and is then reflected in  the Ramond fluxes in supergravity.
We will identify two distinct classes of solutions to the  
supergravity equations, distinguished by the negative or positive 
 sign of $\phi$ at the end of space, which we associate with $n=2,4$, respectively
(see also the discussion at the end of Section 2 in Ref.~\cite{Pilch:2000ue}).
We display  the supersymmetric solutions and  summarise  their known properties 
 in Appendix~\ref{Sec:susy}. 

We reconsider the system consisting of the scalar $\phi$ coupled 
 to gravity in $D=5$ dimensions, and describe more general classes of solutions
 with respect to the literature.
These more general deformations break
explicit supersymmetry and scale invariance, hence lifting the space of vacua,
and modifying the spectrum of the theory.
We focus on solutions
that involve either of two possibilities.
\begin{itemize}
\item In Sects.~\ref{Sec:negativeDWSolutions} and~\ref{Sec:positiveDWSolutions}, we display 
singular domain wall solutions  that generalise the supersymmetric
ones while preserving Poincar\'e invariance in four dimensions.
The dual field theory is deformed by  mass terms, 
breaking supersymmetry,
R-symmetry, and scale invariance.
We compute the spectrum of 
fluctuations---which barring the singularity would be interpreted as
 bound states of the dual field theory in four dimensions---and the behaviour of the quark-antiquark potential 
between static sources, generalising  the results of Ref.~\cite{Brandhuber:1999jr}. 
We discover  one  new special subclass of mildly singular solutions, that
yield a long-distance potential $E_W\propto 1/L^{2}$ persisting up to infinite  separation $L$.
\item In Sect.~\ref{Sec:Confining}, we identify background solutions for which one of the dimensions of the external space-time
is compactified on a circle,  which shrinks smoothly to zero size at some finite value of the radial (holographic) direction.
These  are regular solutions, and the dual field theory
yields linear confinement in three dimensions, as explicitly shown by
 the Wilson loops. We compute the spectrum of fluctuations,  generalising the results of Ref.~\cite{Brower:2000rp}, 
 and discover new features, such as the emergence of an approximate dilaton.
\end{itemize}

\subsection{Sigma-model in $D=5$ dimensions}
\label{Sec:Action}

We denote with hatted symbols quantities characterising the theory in 
$D=5$ dimensions.
The action of the canonically normalised scalar ${\phi}$ 
coupled to gravity is the following (in the notation of Ref.~\cite{Elander:2020csd}):
\beqs
{\cal S}_5 =\int\di^5x \sqrt{-\hat{g}_5}\left(\frac{\hat{\cal R}_5}{4}\,
-\,\frac{1}{2}\hat{g}^{\hat{M}\hat{N}}\partial_{\hat{M}}
{\phi}\partial_{\hat{N}}{\phi} \,-\,{\cal V}_5\right)\,.
\label{Eq:ActionD}
\eeqs
Here $\hat{g}_5$ is the determinant of the metric, $\hat{g}^{\hat{M}\hat{N}}$ its inverse,
and ${\cal \hat{R}}_5$ the Ricci scalar, while ${\cal V}_5$ is the potential.

The Domain Wall (DW) solutions manifestly preserve Poincar\'e invariance in four dimensions.  They can be obtained 
by adopting the following ansatz for the metric:
\beqs
\di s_{DW}^2&=&e^{2{\cal A}(\rho)}\di x^2_{1,3}\,+\,\di \rho^2\,.
\label{Eq:ansatzD}
\eeqs
By assumption,  the only non-trivial functions 
determining the background
are ${\cal A}(\r)$ and $\phi(\r)$, with no dependence on other coordinates.
The resulting second-order equations of motion are the following:
\beqs
\label{Eq:eq1}
0&=&\partial_\r^2 \phi+4\partial_{\r}{\cal A}\partial_\r \phi -\partial_{\phi} {\cal V}_{5}\,,\\
\label{Eq:eq2}
0&=&4(\partial_\r{\cal A})^2 +\partial_\r^2 {\cal A} +\frac{4}{3} {\cal V}_5 \,,\\
\label{Eq:eq3}
0&=& 6 (\partial_\r {\cal A})^2 -  \partial_\r\phi \partial_\r \phi +2 {\cal V}_5\,.
\eeqs

The conventions we are using~\cite{Berg:2005pd} in writing the action in Eq.~(\ref{Eq:ActionD}) are such that \emph{if}
the potential ${\cal V}_5$ of the model can be written in terms of a 
superpotential ${\cal W}$ satisfying
\beqs
{\cal V}_5&=&\frac{1}{2}(\partial_{\phi}{\cal W})^2\,-\,\frac{4}{3}{\cal W}^2\,,
\label{Eq:super}
\eeqs
for the metric ansatz $\di s_{DW}^2$,
\emph{then}  the solutions to the first-order equations 
\beqs
\partial_{\rho}{\cal A}&=&-\frac{2}{3}{\cal W}\, \label{Eq:FirstOrderA}
~~~~{\rm and}~~~~\partial_{\rho}\phi\,=\,\partial_{\phi}{\cal W}\,,
\eeqs
are also solutions to the second-order Eqs.~(\ref{Eq:eq1})-(\ref{Eq:eq3}).

\begin{figure}[t]
	\centering
	\begin{picture}(200,130)
\put(0,0){\includegraphics[width=.37\textwidth]{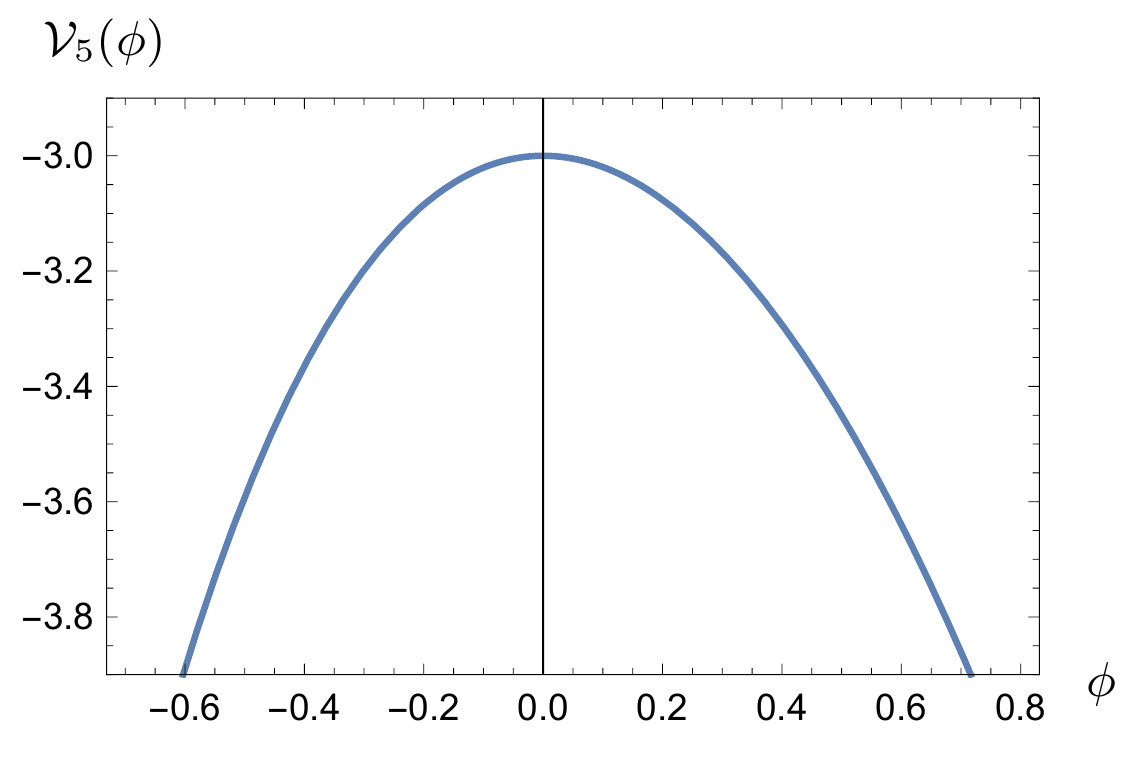}}
	\end{picture}
	\vspace{0mm}
	\caption{The potential ${\cal V}_5$ in Eq.~(\ref{Eq:potential5}) for
	 the sigma-model in $D=5$ dimensions, as a function of
	the scalar $\phi$.}
	\label{Fig:potential}
\end{figure}

The  superpotential is the following~\cite{Bianchi:2000sm,Freedman:1999gk,Brandhuber:1999hb,Bianchi:2001kw}:
\beqs
{\cal W}&=&-e^{-\frac{2\phi}{\sqrt{6}}}-\frac{1}{2}e^{\frac{4\phi}{\sqrt{6}}}\,\,,
\eeqs
and admits an exact AdS$_5$ solution with unit scale.
The potential is given by
\beqs
{\cal V}_5(\phi)&=&-e^{-\frac{4\phi}{\sqrt{6}}}-{2}e^{\frac{2\phi}{\sqrt{6}}}\,,
\label{Eq:potential5}
\eeqs
and is depicted in Fig.~\ref{Fig:potential}.

The first-order equations admit solutions that yield a departure from AdS$_5$ in the interior of the geometry,
which may correspond to $n=2$
(D3 branes distributed on ${\cal B}^2$)
or $n=4$ 
(D3 branes  on ${\cal B}^4$).
For small $\phi$ one finds that ${\cal W}\,\simeq\,-\frac{3}{2}-\phi^2\,+\,\cdots$,
hence these solutions are interpreted in terms the VEV of an operator of dimension 
$\Delta=2$ in the dual field theory.
This saturates the BF bound and, with respect to  Refs.~\cite{Elander:2020ial,Elander:2020fmv},
 brings this study in closer contact with the arguments in Ref.~\cite{Kaplan:2009kr}.

\subsection{Reduction to $D=4$ dimensions}
\label{Sec:Reduction}
We want to model a confining dual field theory in 
	three dimensions. Following Ref.~\cite{Witten:1998zw}, we therefore assume that one 
	spatial dimension is a circle, the size of which
	may depend on the radial direction parametrised by $\rho$ in the five-dimensional geometry, and we hence allow for the breaking of four-dimensional Poincar\'e invariance.\\
	\indent Regular background solutions in which the size of 
	the circle shrinks smoothly to zero (at some finite value of the radial direction $\r=\r_o$)
	introduce a mass gap in the (lower-dimensional) dual field theory,
	and exhibit the physics of confinement---we 
	discuss how  in Sects.~\ref{Sec:positiveDWSolutions} and~\ref{Sec:Confining}.\\ 
	\indent We elect to describe the geometry by applying dimensional reduction of the gravity
	theory to four dimensions, with the introduction of a new dynamical scalar that encodes the size of the circle. In the remainder of this subsection we provide the 
	technical details of the construction.\\
\indent We reduce  to $D=4$ dimensions by adopting the following ansatz,
as in Refs.~\cite{Elander:2020csd} and~\cite{Elander:2013jqa}:
\beqs
\di s_5^2&=&e^{-2\chi(r)}\di s_4^2 \,+\,e^{4\chi(r)} \di \eta^2\,,
\label{Eq:ds25}
\eeqs
where the angle $0\leq \eta<2\pi$ parametrises a circle,
 the four-dimensional metric takes the domain wall form
\beqs
\di s_4^2 &=&e^{2A(r)}\di x_{1,2}^2 \,+\,\di r^2\,,
\eeqs
and the new background scalar $\chi(r)$ and warp factor $A(r)$
depend only on the new radial coordinate $r$. 

The action  in $D=4$ dimensions is
\beqs
{\cal S}_4\hspace{-1pt}=\hspace{-3pt}\int\hspace{-2pt}\di^4 x\sqrt{-{g}_4}\hspace{-2pt}\left[\hspace{-1pt}\frac{{\cal R}_4}{4}
-\frac{{g}^{MN}}{2}{G}_{ab}\partial_{{M}}
{\Phi}^a\partial_{{N}}{\Phi}^b -{\cal V}_4\hspace{-1pt}\right]\hspace{-1pt},
\label{Eq:Action4}
\eeqs
where the sigma-model metric for the scalar fields ${\Phi}^a=\{\phi,  \chi\}$ is $G_{ab}={\rm diag}\left(1, 3\right)$,
and the potential is 
\beqs
{\cal V}_4&=&e^{-2 \chi}{\cal V}_5\,.
\eeqs
We explicitly verified that
$
{\cal S}_5=\int \di \eta \left({\cal S}_4\frac{}{}+\frac{}{}\partial{\cal S}\right)
$,
where
\beqs
\partial {\cal S}&=&\int\di^4x\partial_M\left(\frac{1}{2}\sqrt{-g_4}g^{MN}\partial_N \chi \right)\,.
\eeqs

After the change of variables $\partial_r=e^{- \chi}\partial_{\rho}$,
the equations of motion for the background read as follows:
\beqs
0=
 \partial_{\r}\phi  \hspace{-1pt}\left(3 \partial_{\r}A \hspace{-1pt}-\hspace{-1pt}\partial_{\r}\chi \right)\hspace{-2pt}+ \partial_{\r}^2\phi +\hspace{-2pt}\sqrt{\frac{8}{3}}
   e^{- \sqrt{\frac{8}{3}} \phi  } \hspace{-2pt}\left[e^{\sqrt{6} \phi  }\hspace{-1pt}-\hspace{-1pt}1\hspace{-1pt}\right]\hspace{-1pt}
      \label{Eq:1}
   ,\\
   0=
   3 \partial_{\r}A  \partial_{\r}\chi - \partial_{\r}\chi ^2\hspace{-2pt}+ \partial_{\r}^2\chi 
   -\frac{2}{3} e^{-\sqrt{\frac{8}{3}} \phi  }
  \hspace{-2pt} \left[2 e^{\sqrt{6} \phi  }+1\hspace{-1pt}\right]\hspace{-1pt}
      \label{Eq:2}
   ,\\
   0=\hspace{-2pt}
   \partial_{\r}A  \partial_{\r}\chi -3 \partial_{\r}A ^2\hspace{-2pt}-\partial_{\r}^2A 
   +2 e^{- \sqrt{\frac{8}{3}} \phi  } \hspace{-2pt}\left[2
   e^{\sqrt{6} \phi  }+1\hspace{-1pt}\right]\hspace{-1pt}
      \label{Eq:3},\\
   0=
   3 \partial_{\r}A ^2-3 \partial_{\r}\chi ^2\hspace{-2pt}-\partial_{\r}
   \phi ^2-2 e^{-\sqrt{\frac{8}{3}} \phi  }\hspace{-2pt} \left[2
   e^{\sqrt{6} \phi  }+1\hspace{-1pt}\right]\hspace{-1pt}.
   \label{Eq:4}
\eeqs
By combining Eqs.~(\ref{Eq:2}) and~(\ref{Eq:3}), we obtain
\beqs
\partial_{\rho}\left[e^{3A-\chi}\left(3\partial_{\rho}\chi-\partial_{\rho} A\right)\right]\,=\,0\,,
\label{Eq:conserved}
\eeqs
which defines a conserved quantity along the flow in $\r$.

\subsection{Lift to type IIB in $D=10$ dimensions}
\label{Sec:Lift}

We take the lift to type IIB supergravity in $D=10$ dimensions from
 Ref.~\cite{Pilch:2000ue}.
The dilaton/axion subsystem is trivial 
(see Sect.~3.2 of Ref.~\cite{Pilch:2000ue}), and
there is no distinction between Einstein and string frames.

We parametrise the five-sphere $S^5$ 
in terms of five angles $0\leq\theta \leq \pi/2$,
$0\leq \tilde{\theta}\leq \pi$,
$0\leq \varphi,\tilde{\varphi} <2\pi$, and $0\leq \psi\leq 4\pi$.
The $SU(2)$ left-invariant 2-forms are 
\beqs
\sigma_1&=&\cos\psi \di \tilde{\theta} + \sin \psi \sin \tilde{\theta} \di \tilde{\varphi}\,,\\
\sigma_2&=&\sin\psi \di \tilde{\theta} - \cos \psi \sin \tilde{\theta} \di \tilde{\varphi}\,,\\
\sigma_3&=&\di \psi + \cos \tilde{\theta}\di \tilde{\varphi}\,,
\eeqs
 normalised according to  $\di \sigma_i=\frac{1}{2}\sum_{jk}\epsilon_{ijk}\sigma_j\wedge  
\sigma_k$.\footnote{Compared to Ref.~\cite{Pilch:2000ue}, we have  $\sigma_{i}$(here)$=2\sigma_{i}$({Ref.~\cite{Pilch:2000ue}}).}
We write the metric of the $S^3$ as $\di \Omega_3^2\equiv\frac{1}{4} \sum_{i=1}^3\sigma_i^2$, or
\beqs
\di \Omega_3^2
=\frac{1}{4}\left[\di\tilde{\theta}^2+\sin^2\tilde{\theta}\di\tilde{\varphi}^2
+\left(\di \psi +\frac{}{}\di \tilde{\varphi}\cos\tilde{\theta}\right)^2\right]\,.
\eeqs

We then follow Ref.~\cite{Pilch:2000ue} (see also Ref.~\cite{PremKumar:2010as}),
with the following identifications: we set the two scalars $\{\alpha,\chi\}$ of Ref.~\cite{Pilch:2000ue}
to $\alpha\equiv \phi/\sqrt{6}$ and $\chi=0$ (not to be confused with $\chi$ in this paper), 
we set the coupling $g=2$, and the normalisation 
in Eq.~(3.14) of Ref.~\cite{Pilch:2000ue} to $a^2=2$. 
The metric in $D=10$ dimensions is
\beqs
\di s^2_{10}&=&\Omega^2 \di s_5^2 + \di\tilde{\Omega}_5^2\,,
\eeqs
where $\di s^2_5$ has been introduced in Eq.~(\ref{Eq:ds25}), while
\beqs
\di \tilde{\Omega}_5^2=\frac{X^{1/2}}{\tilde{\r}^3}\left(
\di \theta^2 + \frac{\tilde{\r}^6}{X}\cos^2\theta\di \Omega_3^2+\sin^2\theta \frac{\di \varphi^2}{X}\right)
\,.
\eeqs
The warp factor in the lift depends on $\r$ and $\theta$, because 
\beqs
\Omega^2&\equiv&\frac{X^{1/2}}{\tilde{\r}}\,,
\eeqs
where the functions determining the backgrounds are
\beqs
\tilde{\rho}&\equiv&e^{\phi/\sqrt{6}}
\,,\\
X&\equiv&
\cos^2\theta + \tilde{\r}^6 \sin^2\theta
\,.
\eeqs

For $\phi=0$ one has $\tilde{\r}=1=X$, and recovers the round $S^5$.
The isometries associated with $\di \Omega_3^2$ and $\di \varphi^2$
match the $SO(4)$ and $SO(2)$ symmetries of the field theory.

By making use of 
the equations of motions for the scalars $\phi$ and $\chi$ and for the function $A$, we find that
$
{\cal R}_{10}=0
$ identically.
Yet, other invariants, such as the square of the Ricci and Riemann tensors, are non-trivial.

\subsection{Rectangular Wilson loops}
\label{Sec:Wilson}

The expectation value of rectangular  Wilson loops of sizes $L$ and $T$ in space and time, respectively,
 is computed 
using the standard holographic prescription~\cite{Rey:1998ik,Maldacena:1998im}
(see also Refs.~\cite{Kinar:1998vq,Brandhuber:1999jr,Avramis:2006nv}).
Open strings, with extrema 
bound to the contour of the loop
on the  boundary  of the space at $\r=+\infty$,
explore the geometry down to the turning point $\hat{\rho}_o$
in the holographic direction,
and the problem reduces to a minimal surface one.
The warp factor $\Omega^2$ 
depends on $\theta$, but we restrict attention
to configurations with $\theta$ held fixed,
and focus on the limiting cases $\theta=0$ and $\theta=\pi/2$.
Taking $T\rightarrow + \infty$, we obtain the effective potential between  static quarks
as a function of the separation $L$ between end-points of the string.

The calculation of the Wilson loop can proceed 
along the lines of the prescription in 
Refs.~\cite{Nunez:2009da,Faedo:2013ota,Kinar:1998vq,Brandhuber:1999jr,Avramis:2006nv}.
Starting from the elements of the metric in $D=10$ dimensions, 
$\di s^2=g_{tt} \di t^2+g_{xx} \di x^2+g_{\r\r} \di \r^2+\cdots$,
we introduce the functions $F^2(\r,\theta)\equiv -g_{tt}g_{xx}$ and $G^2(\r,\theta)\equiv 
-g_{tt}g_{\r\r}$, and
the convenient quantity
\beqs
V^{2}_{\rm eff} (\r,\hat{\rho}_o)\equiv
\frac{F^2(\r)}{F^2(\hat{\r}_o)G^2(\r)}\left(F^2(\r)-F^2(\hat{\r}_o)\right)\,,
\eeqs
where the dependence on (constant) $\theta$ is implicit.
The separation between the end points of the string is 
\beqs
L(\hat{\rho}_o)&=&2\int_{\hat{\r}_o}^{\infty} \di {\rho} \frac{1}{V_{\rm eff} (\r,\hat{\rho}_o)}\,,
\eeqs
and the profile of the string in the $(\r,x)$-plane is
\beqs
x(\r)&=&\left\{
\begin{array}{cc}
\int_{\r}^{\infty} \frac{\di y}{V_{\rm eff}(y,\hat{\r}_o)}\,, & x<L(\hat{\rho}_o)/2\,,\\
L(\hat{\rho}_o)-\int_{\r}^{\infty} \frac{\di y}{V_{\rm eff}(y,\hat{\r}_o)}\,, & x>L(\hat{\rho}_o)/2\,.
\end{array}
\right.
\eeqs

The energy of the resulting configuration is 
\beqs
E(\hat{\r}_o)&=&2\kappa\int_{\hat{\r}_o}^{\infty} \di {\rho} 
\sqrt{\frac{F^2(\r)G^2(\r)}{F^2(\r)-F^2(\hat{\rho}_o)}}\,.
\label{Eq:stringenergy}
\eeqs
As  $g_{xx}=-g_{tt}=\Omega^2e^{2A-2\chi}$
and $g_{\r\r}=\Omega^2$, we find the $\theta$-dependent functions $F^2(\r,\theta)=X\tilde{\rho}^{-2}e^{4A-4\chi}$  
and $G^2(\r,\theta)=X\tilde{\rho}^{-2}e^{2A-2\chi}$, also written explicitly as
\beqs
F^2(\r,\theta)=(\cos^2\theta + e^{\sqrt{6}\phi} \sin^2\theta)e^{4A-4\chi-2\phi/\sqrt{6}}\,,\\
G^2(\r,\theta)=(\cos^2\theta + e^{\sqrt{6}\phi} \sin^2\theta)e^{2A-2\chi-2\phi/\sqrt{6}}\,.
\eeqs
Eq.~(\ref{Eq:stringenergy}) is UV-divergent, requiring the
introduction of $\r_{\Lambda}$ as a UV cutoff, and to define the 
regulated $E_{\Lambda}(\hat{\r}_o)$ by restricting the range of integration.
We define the following:
\beqs
\Delta E_{\Lambda,\theta}&\equiv&2\kappa\int_{{\r}_o}^{\rho_{\Lambda}} \di {\rho} 
G(\r,\theta)\,,
\label{Eq:counterterm}
\eeqs
where the integral extends all the way to the end of space $\r_o$,
choose the case $\theta=0$ as a counterterm,
and finally define the renormalised energy as
\beqs
E_{W}(\hat{\rho}_o)&\equiv&\lim_{\rho_{\Lambda}\rightarrow +\infty}
\left(E_{\Lambda}(\hat{\r}_o)\frac{}{}-\Delta E_{\Lambda,0}\right)\,.
\eeqs

In confining theories,  at large separations $L(\hat{\rho}_o)$
the energy grows linearly and the string tension is given by
\beqs
\sigma_{\rm eff}&\equiv&\lim_{\hat{\r}_o\rightarrow \r_o}\frac{\di E_W(\hat{\rho}_o)}{\di L(\hat{\rho}_o)}\,=\,F({\r}_o)\,.
\eeqs
A limiting configuration consists of two straight rods at distance $L$, 
both with fixed $\theta$, extending from the boundary
 to the end of space, connected by a straight portion of string at $\hat{\r}_o=\r_o$.
Its energy is $E_{\theta}=\Delta E_{\Lambda,\theta}-\Delta E_{\Lambda,0}
+F(\r_o,\theta)\,L$. If $F(\r_o,\theta)$ vanishes,
this configuration is indistinguishable from two disconnected ones,
yielding screening in the dual theory---barring the caveats discussed in Ref.~\cite{Bak:2007fk}.
We set the normalisation $\kappa=1$ from here on.
There may be cases in which this procedure 
shows the emergence of a phase transition for the theory living on the probe~\cite{Brandhuber:1999jr}
(see also the discussions in Refs.~\cite{Faedo:2013ota,Faedo:2014naa}).

\section{Background solutions}
\label{Sec:Bg}

We classify in this section the background solutions we are interested in.
We present their UV and IR expansions, and discuss curvature invariants and Wilson loops.

\subsection{UV expansions}
\label{Sec:UV}

All the solutions of interest have the same asymptotic UV expansion, 
and they all correspond to deformations of the same dual theory.
We  expand them for $z=e^{-\r}\ll 1$.
\beqs
\phi(z)&=&
z^2 {\phi_2}+z^2 {\phi_{2l}} \log ( z)+\nonumber\\
   &&+\frac{ \sqrt{6} }{12} z^4
   \left(2{\phi_2}^2-4 {\phi_2} {\phi_{2l}}+3
    {\phi_{2l}}^2\right)+\nonumber\\
   &&   +\frac{\sqrt{6}}{3} z^4 \log ( z) 
   \left( {\phi_2} {\phi_{2l}}- {\phi_{2l}}^2\right)+\nonumber\\
   &&   
   +\frac{z^4 {\phi_{2l}}^2 \log^2( z)}{\sqrt{6}}
   +{\cal O}(z^{6})
\,,\\
\chi(z)&=&
\chi_U-\frac{\log (z)}{2}+{\chi_4} z^4-\frac{1}{6} 
z^4 {\phi_2} {\phi_{2l}} \log ( z)+\nonumber\\
   &&   
-\frac{1}{12} z^4 {\phi_{2l}}^2 \log ^2( z)
      +{\cal O}(z^{6})
\,,\\
A(z)&=&
{A_U}-\frac{3 \log (z)}{2}
-\frac{1}{2} z^4 {\phi_2} {\phi_{2l}} \log( z)+\nonumber\\
&&
+
z^4 \left(\frac{{\chi_4}}{3}
-\frac{1}{36} \left(8 {\phi_2}^2+{\phi_{2l}}^2\right)\right)+
\nonumber\\
&& -\frac{1}{4} z^4
   {\phi_{2l}}^2 \log ^2( z) 
+{\cal O}(z^{6})
\,.
\eeqs
The integration constant $A_U$  can be reabsorbed and set to zero,
while $\chi_U$ can be removed by a shift of radial coordinate $\r$.
 $\phi_2$ is associated with the VEV of the aforementioned dimension-2 operator of the dual field theory, 
 and $\phi_{2l}$ with its (supersymmetry-breaking) coupling.
The parameter $\chi_4$ is associated with the VEV of a dimension-4 operator which triggers
confinement.

\begin{figure}[t]
	\centering
	\begin{picture}(330,245)
\put(0,0){\includegraphics[width=.48\textwidth]{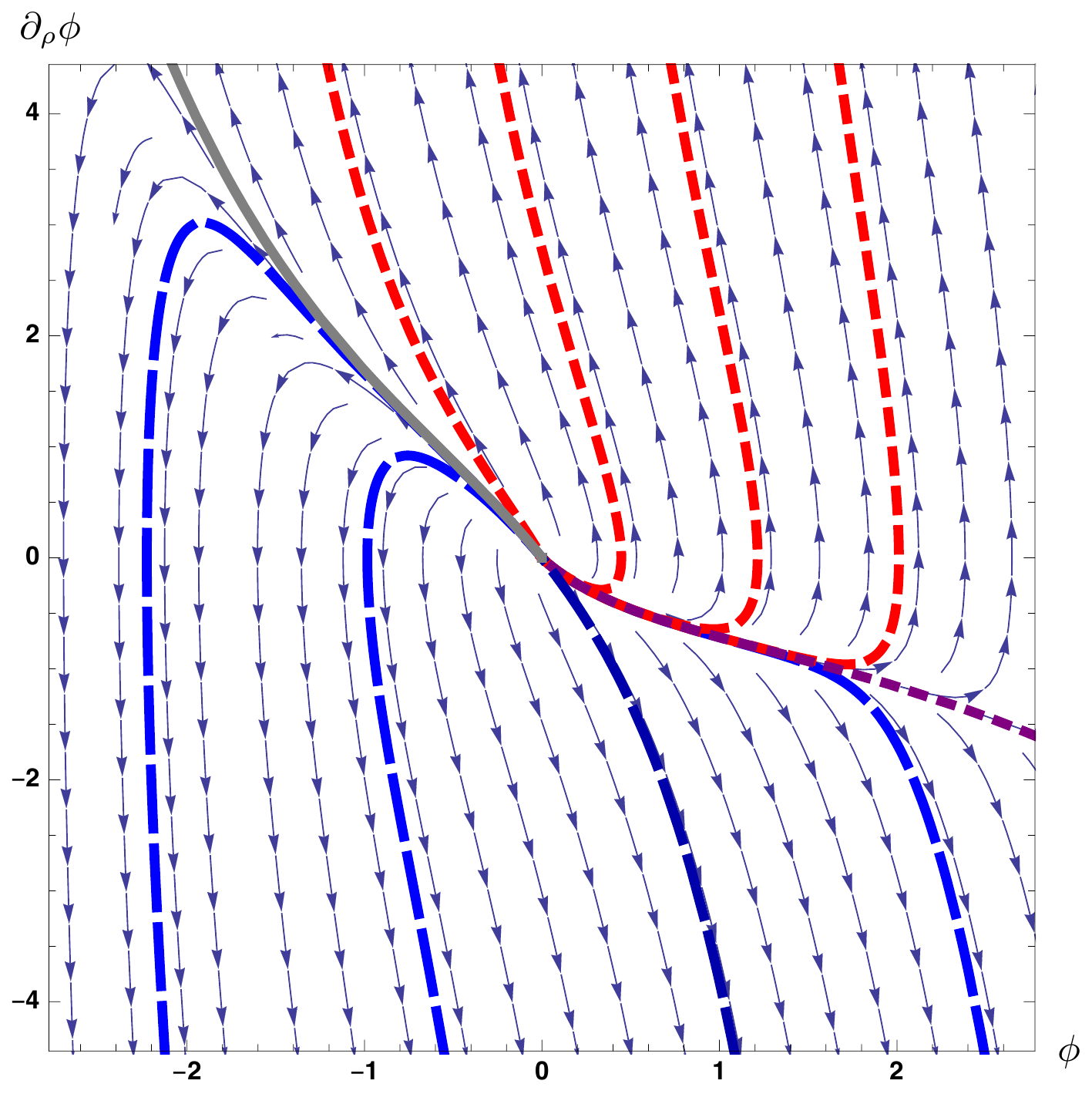}}
\end{picture}
	\vspace{0mm}
	\caption{Stream plot of  the DW solutions
	departing from the trivial fixed point at $(\phi,\partial_{\rho} \phi)=(0,0)$.
	The dashed (red) solutions are examples of the negative DW family described by the IR expansions in Eqs.~(\ref{Eq:phiDWn})
	and~(\ref{Eq:aADWn}). The long-dashed (blue) solutions are examples of the positive DW family with IR expansions in Eqs.~(\ref{Eq:phiDWp})
	and~(\ref{Eq:aADWp}); the case of the supersymmetric solution in Eqs.~(\ref{Eq:phisusy2}) and~(\ref{Eq:aAsusy2}) is denoted by a darker shade of blue.
	The two special solutions are depicted by the continuous thick (grey) line---for the case of the supersymmetric solution in 
	Eqs.~(\ref{Eq:phisusy1}) and~(\ref{Eq:aAsusy1}), or Eqs.~(\ref{Eq:phis2}) 
	and~(\ref{Eq:aAs2})---and the short-dashed (purple) line---for the case described by Eqs.~(\ref{Eq:phis1}) 
	and~(\ref{Eq:aAs1}).
	\label{Fig:stream}}
\end{figure}

Domain wall (DW) solutions in $D=5$ dimensions are recovered (locally) for
${\cal A}=A-\chi=2\chi$, yielding two constraints 
on the five parameters of a generic solution:
\beqs
A_U&=&3\chi_U\,,~~~~~~\chi_4\,=\,-\frac{1}{96}(8\phi_2^2+\phi_{2l}^2)\,.
\label{Eq:DWconditions}
\eeqs
We illustrate the behaviour of the singular DW solutions with the comprehensive catalogue
in Fig.~\ref{Fig:stream}. We devote to them Sects.~\ref{Sec:negativeDWSolutions} and \ref{Sec:positiveDWSolutions}
(and Appendix~\ref{Sec:susy}),
before discussing  in Sect.~\ref{Sec:Confining} the scion of
regular solutions corresponding to confining theories in three dimensions.

\subsection{The negative DW family}
\label{Sec:negativeDWSolutions}

\begin{figure*}[t]
	\centering
	\begin{picture}(410,280)
\put(0,140){\includegraphics[width=.38\textwidth]{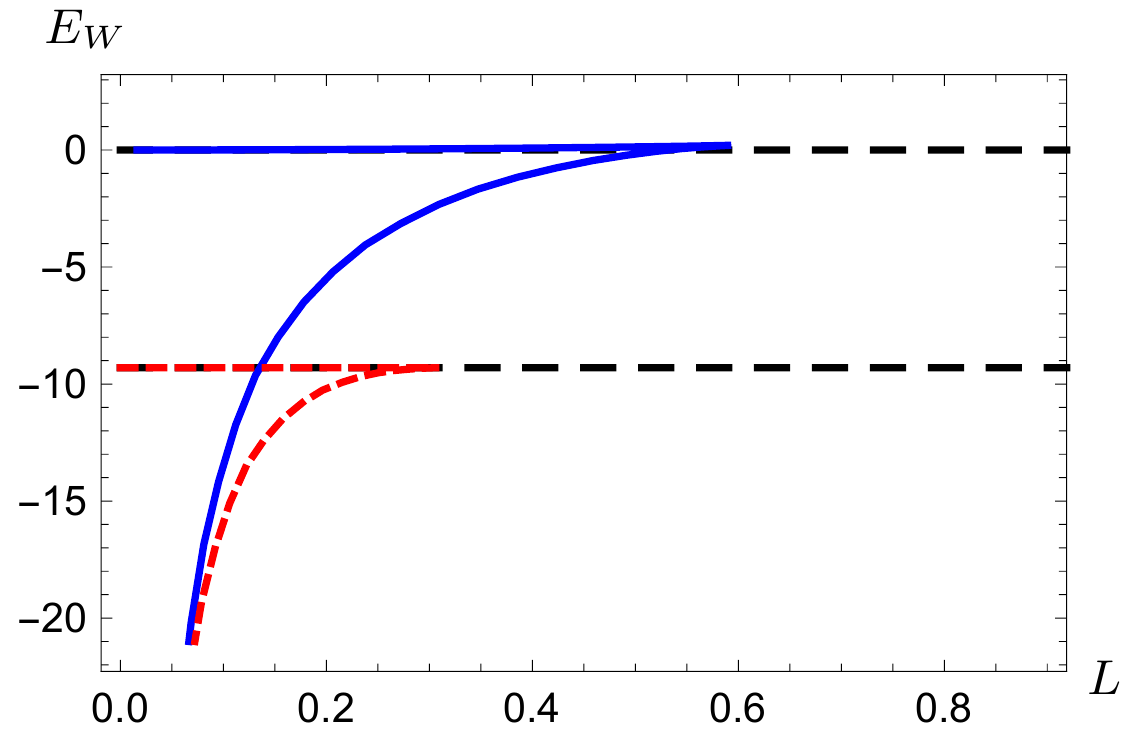}}
\put(209,140){\includegraphics[width=.39\textwidth]{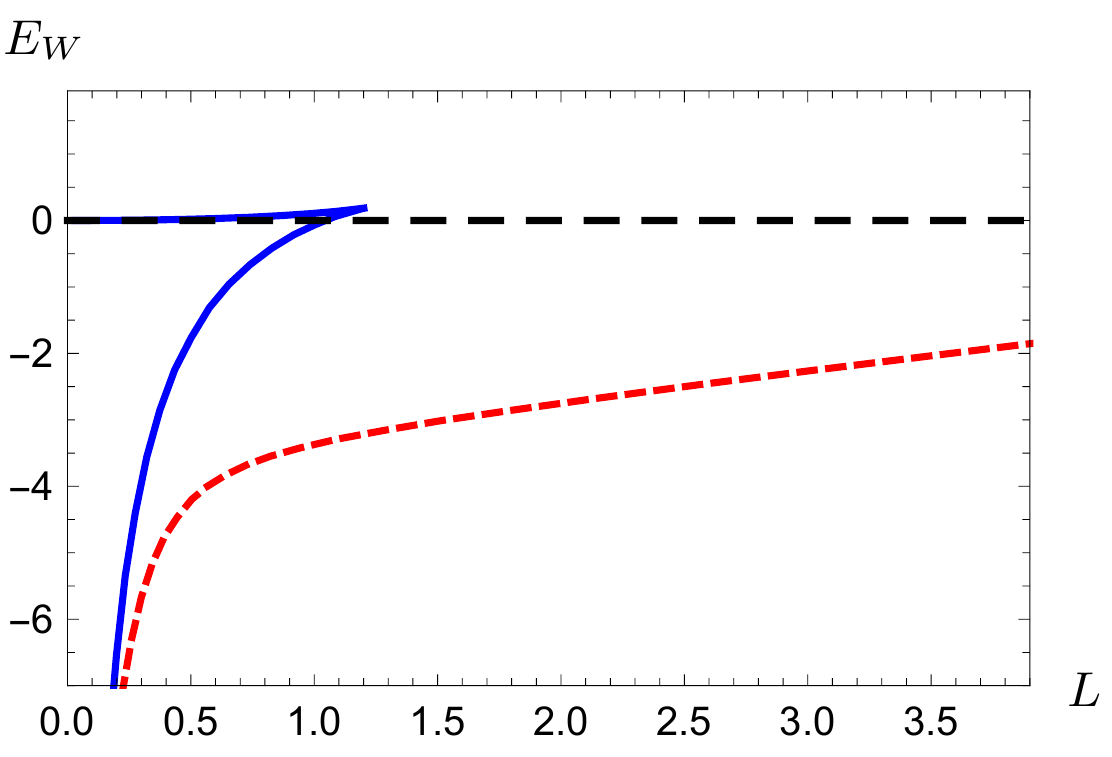}}
\put(0,0){\includegraphics[width=.38\textwidth]{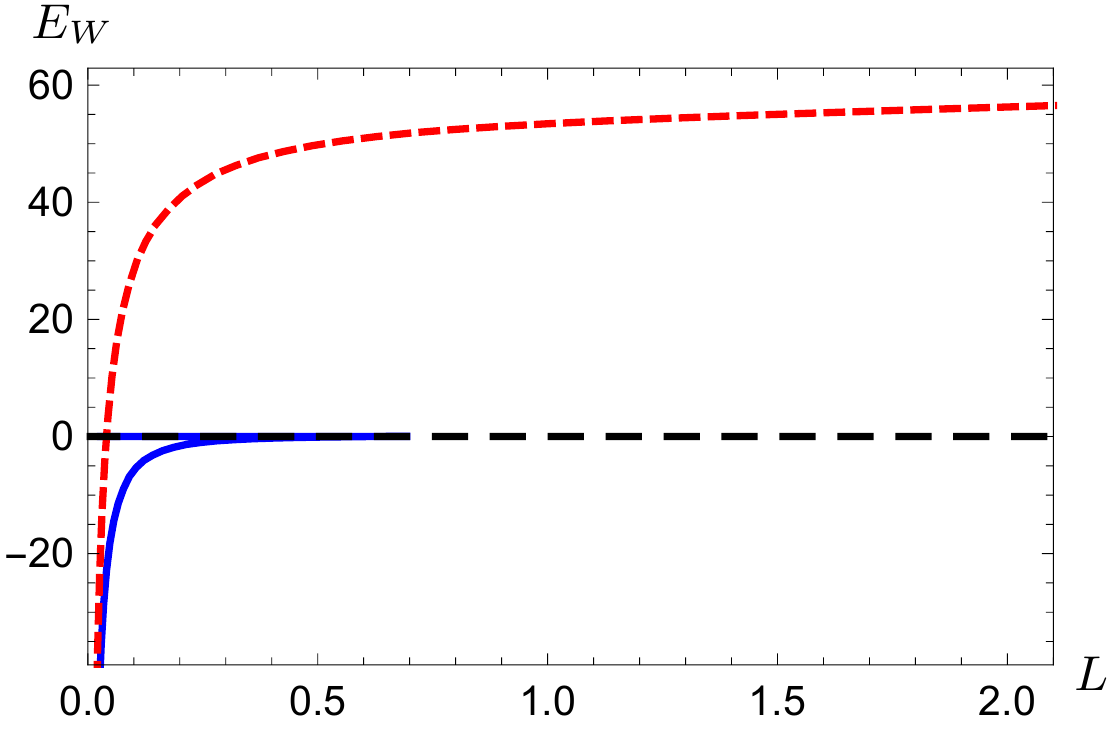}}
\put(200,0){\includegraphics[width=.40\textwidth]{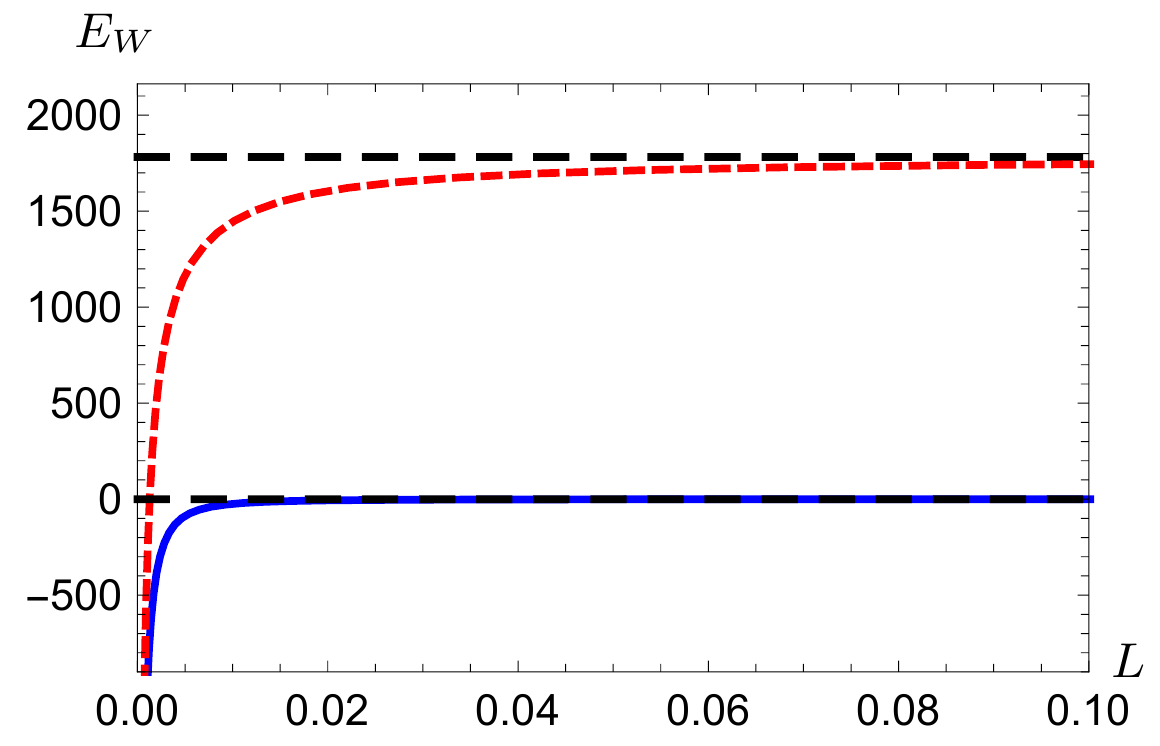}}
	\end{picture}
	\vspace{0mm}
	\caption{Energy $E_W$ as a function of the separation $L$ in space,
	computed with rectangular Wilson loops by applying the  prescription of gauge-gravity dualities, 
	for four different backgrounds
	belonging to classes  of non-supersymmetric DW solutions chosen to have ${\cal A}_I=0$.
	Left to right, and top to bottom, the panels show the results for the following backgrounds: 
	the $(\phi_-,{\cal A}_-)$ solution with $\phi_o=-1$,
	the $(\phi_+,{\cal A}_+)$ solution with $\phi_o=-1$,
	the $(\phi_+,{\cal A}_+)$ solution with $\phi_o=+1$,
	and the special  $(\phi_\infty,{\cal A}_\infty)$ solution corresponding to $\phi_o\rightarrow+\infty$.
	The horizontal long-dashed (black) lines denote  string configurations sitting at the end of space,
	the  continuous (blue) lines depict configurations with $\theta=0$, and the short-dashed (red) lines represent
	 configurations
	with $\theta=\pi/2$. 
	\label{Fig:WilsonDW}}
\end{figure*}

The DW solutions  for which $\phi$ diverges to $-\infty$ 
at the end of space generalise the $n=2$ supersymmetric case---see
 the dashed (red) lines in Fig.~\ref{Fig:stream}.
The background functions, evaluated near the end of space $\r_o$,  are
\beqs
\phi_-(\r)&=&{\phi_o}+\frac{1}{2}   \sqrt{\frac{3}{2}} \log (\r-\r_o)
+\nonumber\\ &&
+\frac{4}{9}   \sqrt{\frac{2}{3}}(\r-\r_o)^2 e^{-4 \sqrt{\frac{2}{3}}{\phi_o}}
+\cdots\,,
\label{Eq:phiDWn}\\
{\cal A}_-(\r)&=&
{\cal A}_I+\frac{\log (\r-\r_o)}{4}
+\nonumber\\ && \label{Eq:aADWn}
+\frac{2}{3} (\r-\r_o) e^{-2   \sqrt{\frac{2}{3}}{\phi_o}}
+\cdots\,.
\eeqs
Ignoring the inconsequential constants ${\cal A}_I$ and $\r_o$,
this one-parameter family of solutions is labelled by the free parameter $\phi_o$. 
The  curvature invariants of the gravity formulation in $D=5$ dimensions
diverge; lifting to $D=10$ dimensions, the singular behaviour first appears in
\beqs
{\cal R}_{10,\hat{M}\hat{N}}{\cal R}_{10}^{\,\,\,\,\,\hat{M}\hat{N}}&=&
\frac{10 e^{-\sqrt{6} \phi_o}}{\cos^2(\theta)(\r-\r_o)^{3/2}}
+\cdots\,,
\eeqs

A special limiting case (corresponding to $\phi_o\rightarrow -\infty$) of the (singular) negative DW solutions is
represented by the thick (grey) line in Fig.~\ref{Fig:stream}, and is
given by the  IR expansions in Eqs.~(\ref{Eq:phis2}) and ~(\ref{Eq:aAs2}).
It satisfies the first-order equations, 
as it  coincides with the supersymmetric solutions $(\phi_2,{\cal A}_2)$ described by
Eqs.~(\ref{Eq:phisusy1}) and~(\ref{Eq:aAsusy1})---the solution corresponding
to the ($n=2$) case of  D3 branes distributed on a disk (${\cal B}^2$).

The study of the Wilson loops 
is exemplified in Fig.~\ref{Fig:WilsonDW}. The top-left panel depicts the case
of backgrounds $(\phi_-,{\cal A}_-)$  with $\phi_o=-1$. The string is tensionless at the end of space, as
$\lim_{\r\rightarrow \r_o}F^2(\r)=0$ for both choices  $\theta=0,\pi/2$.
The separation $L$ converges to zero for strings with end points at $\r=+\infty$,
when the turning point of the string configuration reaches the end of space, 
yielding the description of a phase transition 
such that the Wilson loop mimics screening  at  large $L$.

\subsection{The positive  DW family}
\label{Sec:positiveDWSolutions}

 In the 
stream plot in Fig.~\ref{Fig:stream},
 the long-dashed (blue) lines are examples of  DW solutions in which 
the scalar $\phi$ diverges to $\phi\rightarrow +\infty$ at the end of space.
Their IR expansions are
\beqs
\phi_+(\r)&=&
{\phi_o}
-\frac{1}{2} \sqrt{\frac{3}{2}} \log (\r-\r_o)
+\nonumber\\ &&
+\frac{8}{15} \sqrt{\frac{2}{3}} e^{\sqrt{\frac{2}{3}}{\phi_o}}(\r-\r_o)^{3/2} 
\label{Eq:phiDWp}
+\cdots\,,\\
{\cal A}_+(\r)&=&
{\cal A}_I
+\frac{\log (\r-\r_o)}{4}
+\nonumber\\ 
 &&
 \label{Eq:aADWp}
 +\frac{32}{45}(\r-\r_o)^{3/2}   e^{\sqrt{\frac{2}{3}}{\phi_o}}
+\cdots\,.
\eeqs

These solutions generalise the supersymmetric one denoted $(\phi_4,{\cal A}_4)$ 
in Eqs.~(\ref{Eq:phisusy2}) and~(\ref{Eq:aAsusy2})---the solutions 
corresponding to the ($n=4$) case of D3 branes distributed on ${\cal B}^4$---to a one-parameter family,
labelled by $\phi_o$.
The supersymmetric case is recovered
with the choice $\phi_o=-\frac{1}{2}\sqrt{\frac{3}{2}}\log\left(\frac{4}{3}\right)$,
and is highlighted by a darker long-dashed line  in Fig.~\ref{Fig:stream}.

The five-dimensional curvature invariants diverge. In $D=10$ dimensions the 
divergence appears in 
\beqs
{\cal R}_{10,\hat{M}\hat{N}}{\cal R}_{10}^{\,\,\,\,\,\hat{M}\hat{N}}&=&
\frac{405 e^{-8\sqrt{\frac{2}{3}}\phi_o}\sin^4(2\theta)}{2048\sin^{10}(\theta)}
+\cdots\,.
\eeqs
The limit $\theta\rightarrow 0$ is singular at the end of space, even in the case of the 
supersymmetric solution---see also Eq.~(\ref{Eq:RicciEquator4}) and the discussion that follows it.

The calculation of the Wilson loops is exemplified in the top-right and bottom-left
panels  in Fig.~\ref{Fig:WilsonDW},  for $\phi_o=-1$ and $\phi_o=+1$, respectively.
For $\theta=0$, once more
$\lim_{\r\rightarrow \r_o}F^2(\r)=0$. The separation $L$ vanishes when 
 the turning point of the string configuration reaches the end of space, 
  as we found for $(\phi_-,{\cal A}_-)$.
But in the case $\theta=\pi/2$,
we find that 
$\lim_{\r\rightarrow \r_o}F^2(\r)=\sigma^2>0$ is finite. The separation $L$ diverges,
and one recovers the linear potential  $E_W \simeq \sigma L $.
When $\phi_o<-\frac{1}{2}\sqrt{\frac{3}{2}}\log\left(\frac{4}{3}\right)$, 
 the assumption of keeping $\theta$ fixed fails, as at small $L$ the configurations with $\theta=\pi/2$
 have lower energy than those with $\theta=0$, while
 at large  $L$ the converse is true.

\subsubsection{Special positive DW solutions}
\label{Sec:specialpositiveDWSolutions}
A limiting case of the positive DW solutions is depicted by the short-dashed (purple) line 
in Fig.~\ref{Fig:stream}. The 
IR expansions are:
\beqs
\phi_{\infty}(\r)&=&\nonumber
\sqrt{\frac{3}{2}}\log\left(\frac{45}{2}\right)
-\sqrt{6}\log(\r-\r_o)
+\\
&&
\label{Eq:phis1}
+\frac{2\sqrt{2}}{59535\sqrt{3}}(\r-\r_o)^6\,+
\cdots\,,\\
{\cal A}_{\infty}(\r)&=&\nonumber
{\cal A}_I+4\log(\r-\r_o)+\\
&&\label{Eq:aAs1}
+\frac{16}{893025}(\r-\r_o)^6\,+
\cdots\,.
\eeqs

The only parameters are the inconsequential ${\cal A}_I$ and $\r_o$.
The five-dimensional curvature invariants diverge, but the lift to $D=10$ dimensions yields
\beqs
{\cal R}_{10}=0=
\lim_{\r\rightarrow \r_o}{\cal R}_{10,\hat{M}\hat{N}}{\cal R}_{10}^{\,\,\,\,\,\hat{M}\hat{N}}\,.
\eeqs
Yet, these solutions are singular as well, 
as illustrated by the simultaneous limits  $\rho\rightarrow \r_o$ 
 and $\theta\rightarrow 0$ of the square of the Riemann tensor:
\beqs
\lim_{\r\rightarrow \r_o}({\cal R}_{10,\hat{M}\hat{N}\hat{R}\hat{S}})^2
=
\frac{9(15+10\sin^2(\theta)+7\sin^4(\theta))}{5\sin^6(\theta)}.
\eeqs

These solutions are the limiting case $\phi_o\rightarrow +\infty$ of the $(\phi_+,{\cal A}_+)$ general class,
and the bottom-right panel in Fig.~\ref{Fig:WilsonDW} shows a peculiarly interesting behaviour
for the quark-antiquark potential.
The separation $L$ is unbounded,
 the potential vanishes for $L\rightarrow + \infty$, and so does the string tension.
We find the potential $E_W\simeq-e^{1/6}/L$ at short $L$,
and  $E_W\simeq-e^{6}/L^2$ at large $L$ (for ${\cal A}_I=0$).

While the results of the study of the Wilson loops for the DW solutions
 are very suggestive, with  the emergence of screening, confining, several types of 
Coulombic potentials and phase transitions,
 they must be all taken with caution;
 all the background solutions discussed so far (and in Appendix~\ref{Sec:susy}) are singular.
Hence, such solutions cannot be considered as complete gravity duals of  field theories, but they provide only approximate 
descriptions that may miss important long-distance details.

\subsection{Confining solutions}
\label{Sec:Confining}

\begin{figure}[t]
	\centering
	\begin{picture}(200,270)
\put(0,135){\includegraphics[width=.38\textwidth]{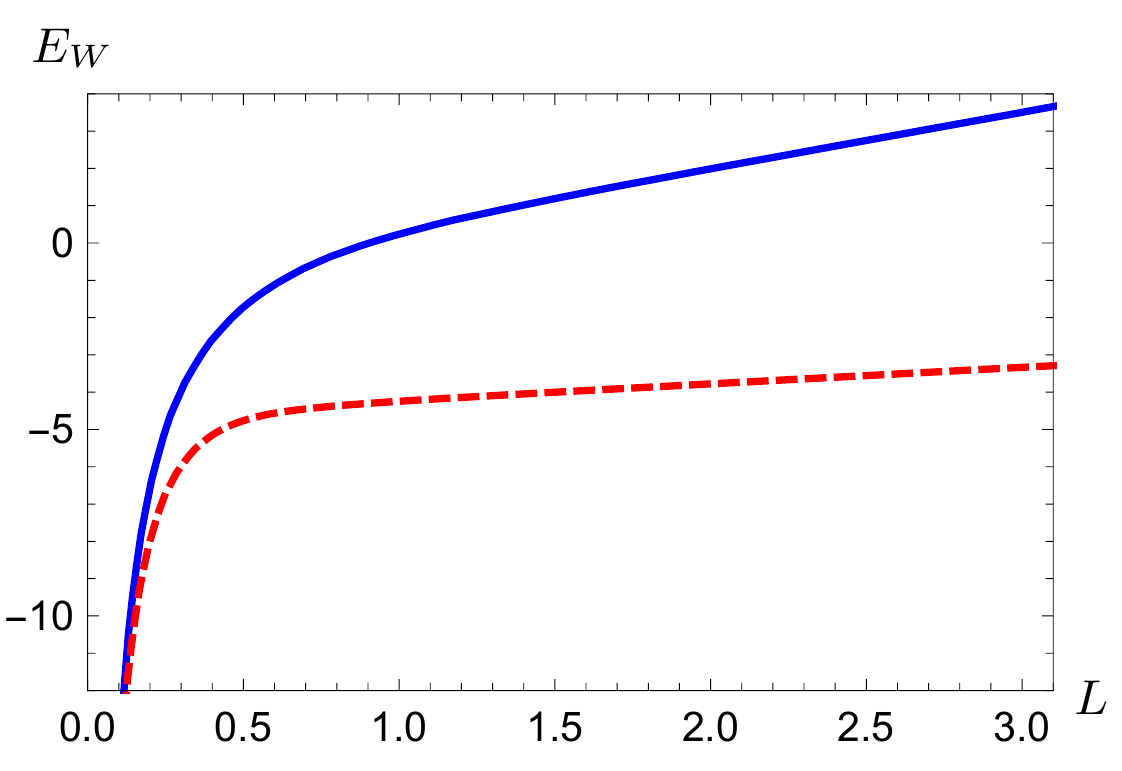}}
\put(0,0){\includegraphics[width=.38\textwidth]{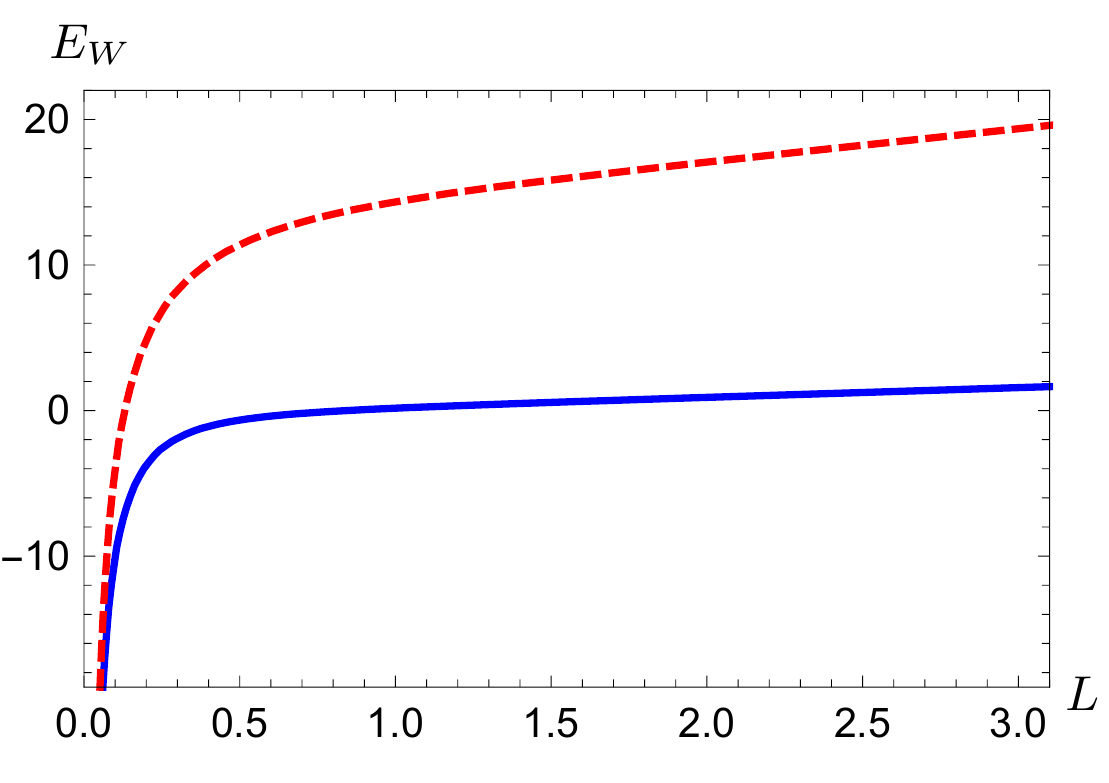}}
	\end{picture}
	\vspace{0mm}
	\caption{Energy $E_W$ as a function of the separation $L$, for two representative confining solutions,
	chosen to have $A_I=0=\chi_I$, and $\phi_I=-1$ 
	 (top) or  $\phi_I=+1$ (bottom).
	Continuous (blue) lines correspond to configurations with $\theta=0$, short-dashed (red) lines to 
	configurations with $\theta=\pi/2$.
	\label{Fig:WilsonCONF}}
\end{figure}  

The solutions of this class are completely regular, and dual
to confining, three-dimensional field theories. 
Here we present their IR expansions, discuss the gravitational invariants, and compute the Wilson loops.
The expansion  in proximity of the
end of space $\r_o$, is
\beqs
\phi_C(\r)&=&
{\phi_I}-\frac{(\r-\r_o)^2 e^{-2 \sqrt{\frac{2}{3}} {\phi_I}} \left(e^{\sqrt{6} {\phi_I}}-1\right)}{\sqrt{6}}+
\nonumber\\ &&
\label{Eq:phiI}
\,+\,{\cal O}\left((\r-\r_o)^4\right)\,,\\
\chi_C(\r)&=&
{\chi_I}+\frac{\log (\r-\r_o)}{2}+
\nonumber\\ &&
-\frac{1}{18} (\r-\r_o)^2   e^{-2 \sqrt{\frac{2}{3}} {\phi_I}} \left(2 e^{\sqrt{6} {\phi_I}}+1\right)+
\nonumber\\ &&
\,+\,{\cal O}\left((\r-\r_o)^4\right)\,,\\
A_C(\r)&=&
{A_I}+\frac{\log (\r-\r_o)}{2}+
\nonumber\\ &&
+\frac{5}{18} (\r-\r_o)^2   e^{-2 \sqrt{\frac{2}{3}} {\phi_I}} \left(2 e^{\sqrt{6} {\phi_I}}+1\right)+
\nonumber\\ &&
\,+\,{\cal O}\left((\r-\r_o)^4\right)\,.
\eeqs
The gravity invariants in five dimensions
are finite, and 
when restricted to the $(\rho,\eta)$ plane, the metric reduces to
\beqs
\di s_2^2&=&\di \rho^2 + e^{4\chi_I}(\r-\r_o)^2 \di \eta^2\,.
\eeqs
We fix $\chi_I=0$ in order to avoid a conical singularity.
The integration constant $A_I$ is trivial and can be reabsorbed by a rescaling of the three
Minkowski directions. The constant $\phi_I$ characterises this one-parameter family of solutions.
The curvature invariants of the lift to $D=10$ dimensions are regular,
for all choices of $0\leq \theta \leq \pi/2$.

In the study of the  rectangular Wilson loop (in  three dimensions) we 
fix $\theta$, and allow the two sides of the 
rectangle to align with time and one non-compact space-like direction.
The static potential $E_W(L)$ is illustrated by Fig.~\ref{Fig:WilsonCONF},
for two representative choices with $\phi_o\pm 1$.
The short-distance Coulombic behaviour gives way to the linear potential
typical of confinement, and $L$ is unbounded. For this reason, with
some abuse of language, we call these regular solutions {\it confining}.
We can compute the string tension, and we find 
\beqs
\sigma_0&\equiv&\lim_{\hat{\r}_o\rightarrow \r_o}F(\hat{\r}_o,0)\,=\,e^{2A_I-2\chi_I-\sqrt{\frac{1}{6}}\phi_I}\,,\\
\sigma_{\pi/2}&\equiv&\lim_{\hat{\r}_o\rightarrow \r_o}F(\hat{\r}_o,\pi/2)\,=\,e^{2A_I-2\chi_I+2\sqrt{\frac{1}{6}}\phi_I}\,.
\eeqs
The configuration with $\theta=0$ has lower energy in the case where $\phi_I>0$, and vice versa.

\section{Mass spectra and probe approximation}
\label{Sec:spectrum}

 The spectrum of small fluctuations of a
  sigma-model coupled to gravity of the form of Eqs.~(\ref{Eq:ActionD}) and~(\ref{Eq:Action4})
   in  generic $D$ dimensions
 can be interpreted in terms of the spectrum of bound states of the strongly-coupled dual field theory,
  by applying the dictionary of gauge-gravity dualities.
We adopt the gauge-invariant formalism described in detail in 
Refs.~\cite{Bianchi:2003ug,Berg:2005pd,Berg:2006xy,Elander:2009bm,Elander:2010wd}.
Due to the divergences in the deep IR and far UV,
we introduce two unphysical  boundaries $\r_1<\r<\r_2$ in the radial direction---the
physical results are recovered in the limits $\r_2\rightarrow +\infty$ and $\r_1\rightarrow \r_o$.
The calculation involves fluctuating solutions 
for which the metric has the DW form in $D$ dimensions.
The confining solutions assume the DW form in
the dimensionally reduced ($D=4$) formulation of the theory. For the confining solutions, it is also understood that in the following equations (\ref{Eq:tensoreom}~-~\ref{Eq:probe}) appearing in this section of the paper, ${\cal A}$ is to be replaced by $A$.

The tensorial fluctuations $\mathfrak e^\mu{}_{\nu}$  are gauge-invariant,  obey the equations of motion
\beqs
\label{Eq:tensoreom}
	\left[ \partial_\r^2 + (D-1) \partial_\r {\cal A} \partial_\r + e^{-2{\cal A}(\r)} M^2 \right] \mathfrak e^\mu{}_\nu = 0 \,,
\eeqs
where  $M$ is the mass in $D-1$ dimensions, and are subject  to Neumann boundary conditions
$
	\left.\frac{}{}\partial_\r \mathfrak e ^\mu{}_{\nu} \right|_{\r_i}= 0$.
The scalar gauge invariant fluctuations $\mathfrak a^a\equiv \varphi^a - \frac{\partial_\r
  \Phi^a}{2(D-2)\partial_\r {\cal A}} h$ are a combination of fluctuations $\varphi^a$ of the  scalar fields
  and $h$ of the trace of the metric. They obey the following equations of motion and boundary conditions
  \begin{widetext}
\beqs
\label{Eq:scalareom}
	0 &=& \Big[ {\cal D}_\r^2 + (D-1) \partial_{\r}{\cal A} {\cal D}_\r + e^{-2{\cal A}} M^2 \Big] \mathfrak{a}^a 
	- \left[  V^a{}_{|c}	+ 
	\frac{4 (\partial_{\r}\Phi^a  V^b +  V^a 
	\partial_{\r}\Phi^b) G_{bc}}{(D-2) \partial_{\r} {\cal A}} + 
	\frac{16  V \partial_{\r}\Phi^a \partial_{\r}\Phi^b G_{bc}}{(D-2)^2 (\partial_{\r}{\cal A})^2} \right] \mathfrak{a}^c\,,\\
\label{Eq:scalarbc}
0&=& \frac{2  e^{2{\cal A}}\partial_{\r} \Phi^a }{(D-2)M^2 \partial_{\r}{\cal A}}
	\left[ \partial_{\r} \Phi^b{\cal D}_\r-\frac{4  V \partial_{\r} \Phi^b}{(D-2) 
	\partial_\r {\cal A}} - V^b \right] \mathfrak a_b + \mathfrak a^a\Big|_{\r_i}\,.
\eeqs
\end{widetext}
The notation 
 follows the conventions of Ref.~\cite{Elander:2010wd}.
The sigma-model metric being  trivial, the covariant derivative simplifies to 
$
V^a_{\,\,\,|c}\equiv D_c V^a\,=\,\partial_c (G^{ab}\partial_b V)\,,
$
and the background-covariant derivative to  ${\cal D}_{\r}\mathfrak{a}^a=\partial_{\r}\mathfrak{a}^a$.

The \emph{probe approximation} is defined
 according to the prescription tested in Ref.~\cite{Elander:2020csd},
and we use it as a diagnostic tool to identify particles coupled
to the trace of the energy momentum tensor, because of their mixing with $h$.
The probe approximation ignores the fluctuation $h$,
in the definition of $\mathfrak a^a$, yielding variables $\mathfrak{p}^a$ 
that  satisfy
\beqs
\label{Eq:probe}
0= \Big[ {\cal D}_\r^2 + (D-1) \partial_{\r}{\cal A} {\cal D}_{\r} - e^{-2A} q^2 \Big] \mathfrak{p}^a-V^a{}_{|c}
	  \mathfrak{p}^c\,,
\eeqs
subject  to Dirichlet boundary conditions
$
 \mathfrak p^a\Big|_{\r_i} = 0$.

The fluctuation $h$ is interpreted as the bulk field coupled to the dilatation operator
in the dual field theory.
If the approximation of ignoring $h$ captures correctly the spectrum, then
the associated scalar particle is not a dilaton. Conversely,
 the probe approximation  either completely misses, or fails to capture the correct mass of, 
 an approximate dilaton.
We tested  these ideas on a large selection of examples  in Ref.~\cite{Elander:2020csd}.

\begin{figure}[t]
	\centering
	\begin{picture}(330,154)
\put(10,0){\includegraphics[width=.45\textwidth]{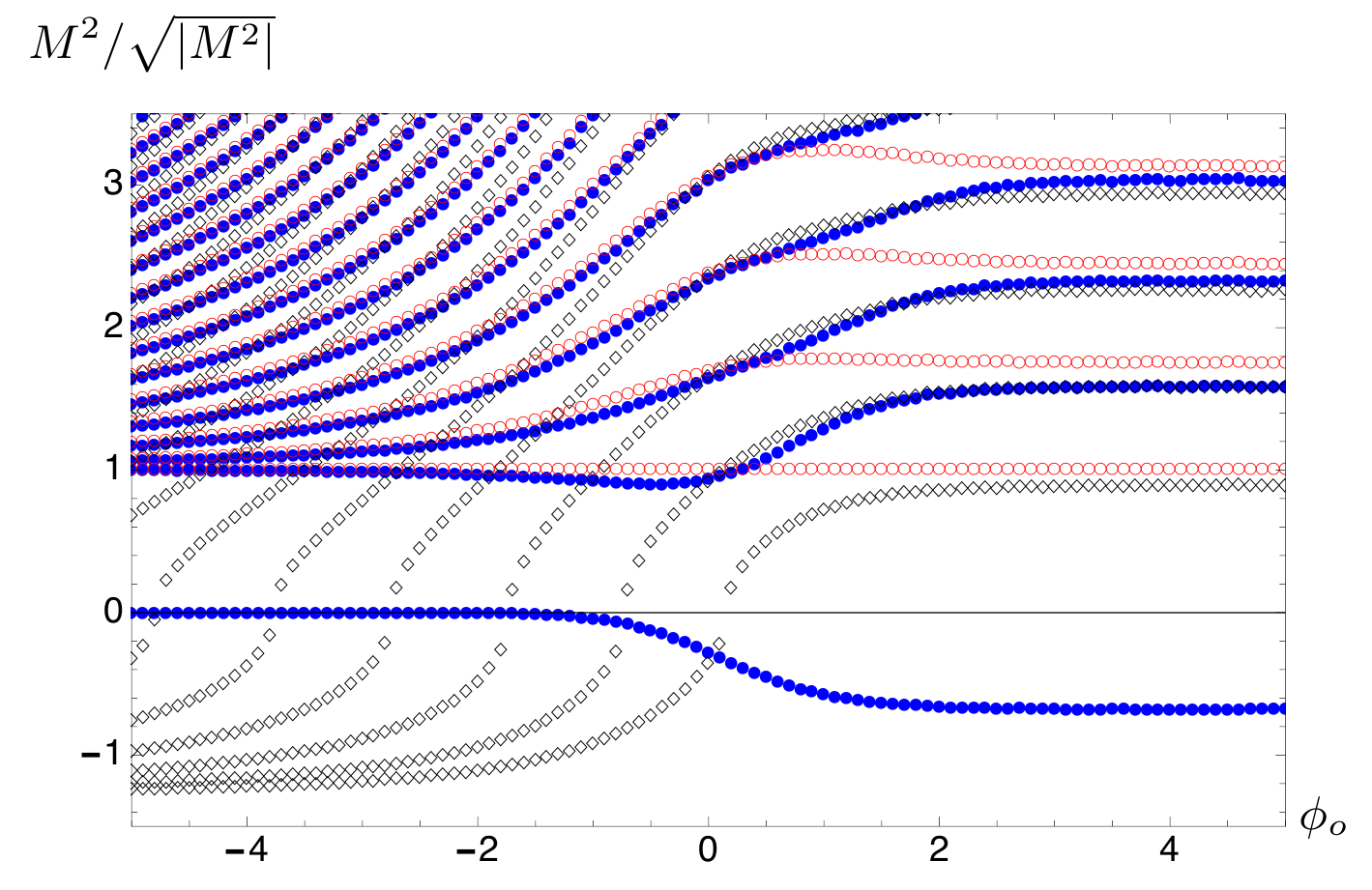}}
	\end{picture}
	\vspace{0mm}
	\caption{Mass spectrum of the fluctuations of the negative DW solutions. The (red) circles are spin-2 states,
	 the (blue) disks are spin-0 states, and the (black) diamonds represent the probe approximation calculation
	 of the same spin-0 masses. The masses $M$ are
	normalised to the lightest spin-2 state, and plotted as a function of 
	$\phi_o$,  defined in Eqs.~(\ref{Eq:phiDWn}) and~(\ref{Eq:aADWn}). For $\phi_o\rightarrow -\infty$
	the backgrounds approach the supersymmetric solution in Eqs.~(\ref{Eq:phisusy1}) 
	and~(\ref{Eq:aAsusy1})---or Eqs.~(\ref{Eq:phis2}) 
	and~(\ref{Eq:aAs2}). The numerical calculations are performed with finite cutoffs $\r_1=10^{-6}$ 
	and $\r_2=8$. We checked explicitly that these choices are close enough to the physical limits 
	$\r_1\rightarrow \r_o=0$ and $\r_2\rightarrow +\infty$ that the numerical results do not display important residual 
	spurious dependence on the cutoffs.
	\label{Fig:spectraDWn}}
\end{figure}

\begin{figure}[t]
	\centering
	\begin{picture}(330,154)
\put(10,0){\includegraphics[width=.45\textwidth]{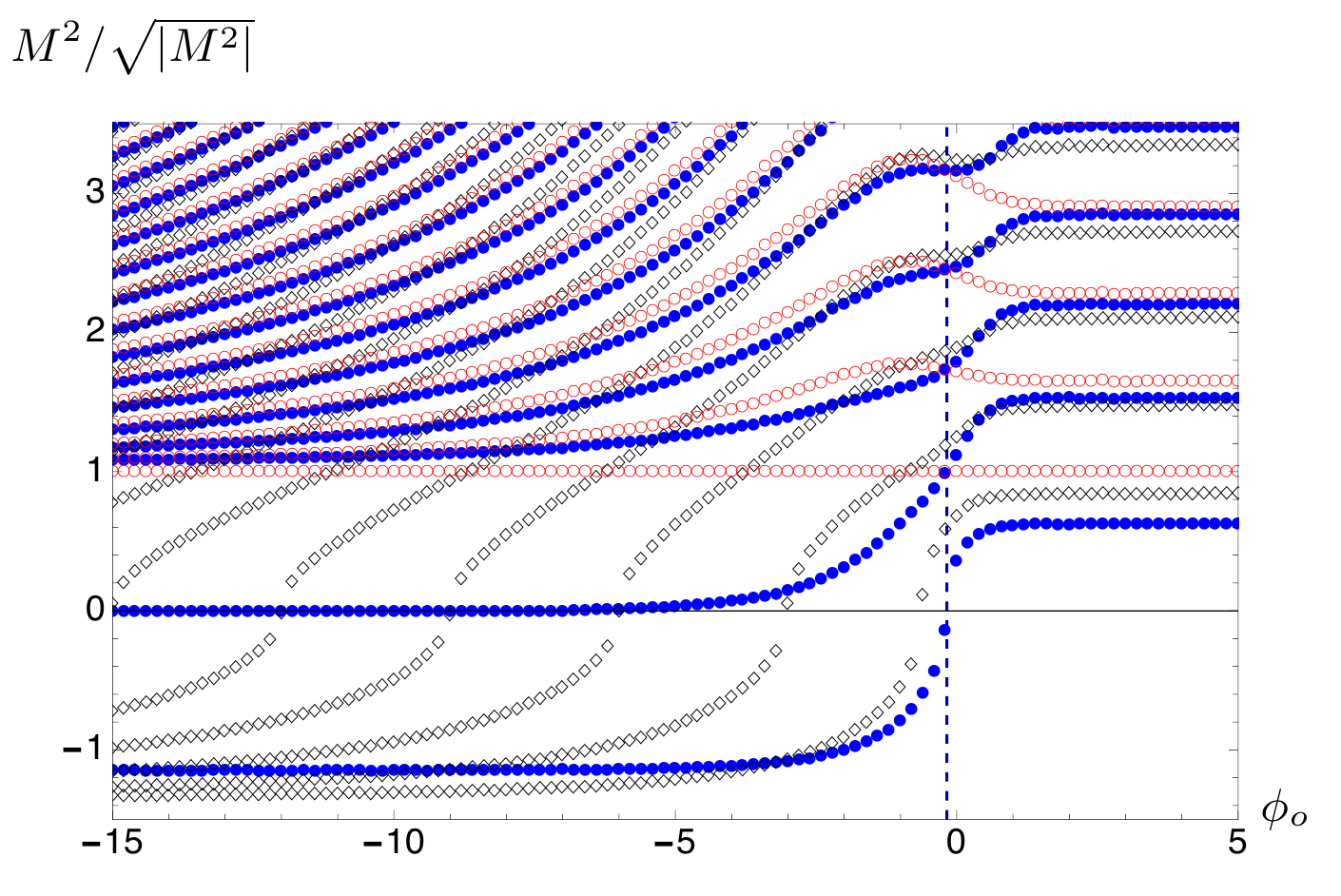}}
	\end{picture}
	\vspace{0mm}
	\caption{Mass spectrum of the fluctuations of the positive DW solutions. The (red) circles are the spin-2 states,
	 the (blue) disks are the spin-0 states, and the (black) diamonds represent the probe approximation calculation
	 of the spin-0 masses. The masses $M$ are
	normalised to  the lightest spin-2 state, and plotted as a function of 
	$\phi_o$,  defined in Eqs.~(\ref{Eq:phiDWp}) and~(\ref{Eq:aADWp}).
The special case of the supersymmetric solutions in Eqs.~(\ref{Eq:phisusy2}) and~(\ref{Eq:aAsusy2}) is 
the choice of $\phi_o$ that yields a massless scalar state, and is marked by a vertical dashed line. The numerical calculations are performed with finite cutoffs $\r_1=10^{-6}$ 
	and $\r_2=8$. We checked explicitly that these choices are close enough to the physical limits 
	$\r_1\rightarrow \r_o=0$ and $\r_2\rightarrow +\infty$ that the numerical results do not display any important  residual
	spurious dependence on the cutoffs.
		\label{Fig:spectraDWp}}
\end{figure}

\begin{figure}[h]
	\centering
	\begin{picture}(330,154)
\put(10,0){\includegraphics[width=.45\textwidth]{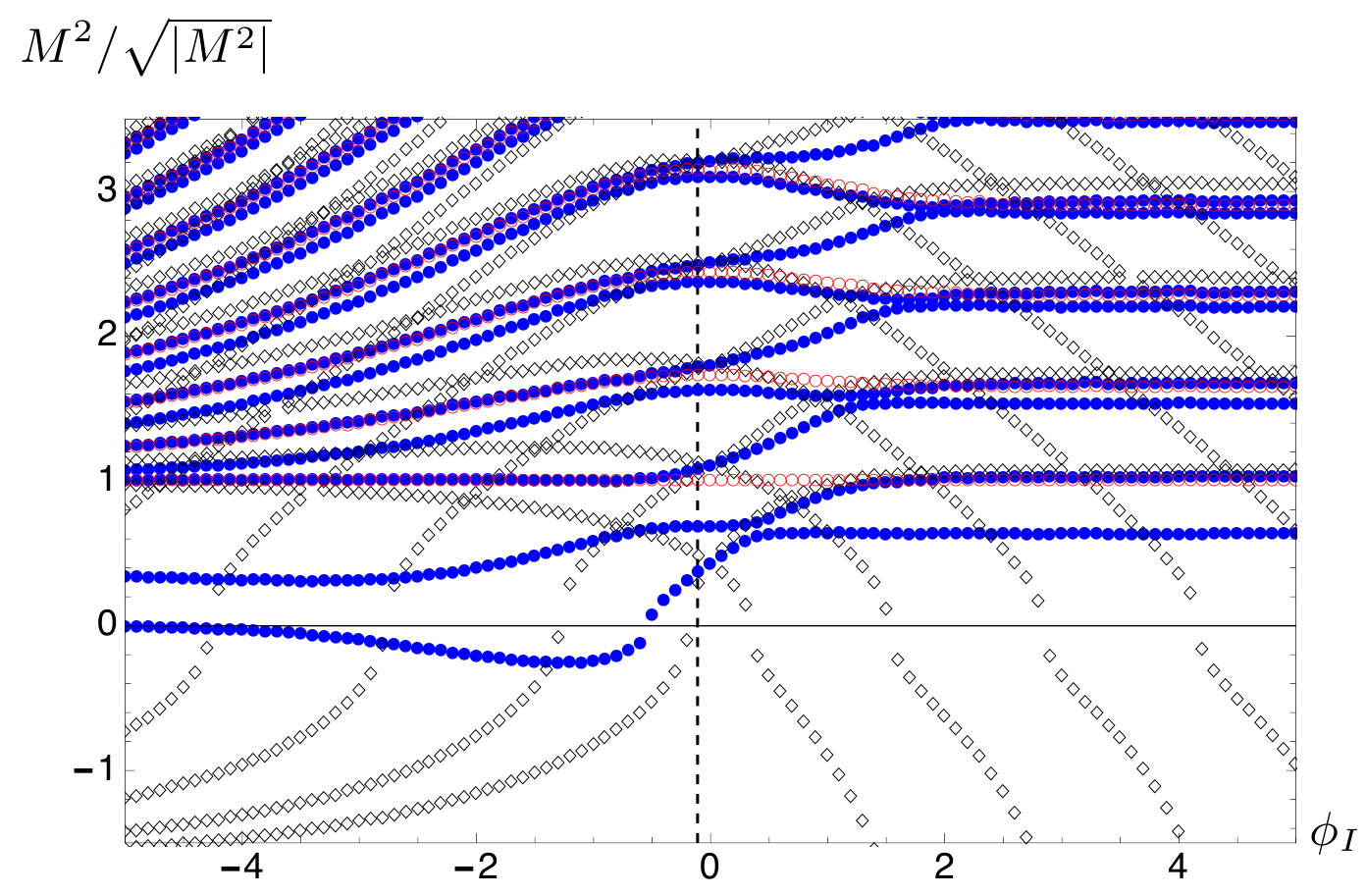}}
	\end{picture}
	\vspace{0mm}
	\caption{Mass spectrum of the fluctuations of the confining solutions. The spectrum can be interpreted in 
	terms of the masses of bound states in a confining theory in three dimensions.
	The (red) circles are the spin-2 states,
	 the (blue) disks are the spin-0 states, and the (black) diamonds represent the probe approximation calculation
	 of the spin-0 masses. The masses $M$ are
	normalised to  the lightest spin-2 state, and plotted as a function of 
	$\phi_I$,  defined in Eq.~(\ref{Eq:phiI}).
	The vertical dashed line denotes the value of the parameter $\phi_I=\phi_I^{\star}\simeq -0.067$ 
	at which a phase transition takes place.
	The numerical calculations are performed with  finite cutoffs $\r_1=10^{-6}$ 
	and $\r_2=8$. We checked explicitly that these choices are close enough to the physical limits 
	$\r_1\rightarrow \r_o=0$ and $\r_2\rightarrow +\infty$ that the numerical results do not display any important  residual 
	spurious dependence on the  cutoffs.
			\label{Fig:spectraConf}}
\end{figure}  

In order to improve the convergence of the numerical computation of the spectrum, 
we make use of the UV expansions for the fluctuations given in Appendix B, 
setting up the boundary conditions such that only the sub-leading modes are retained. 
This is the customary prescription, as well as the one selected by the boundary conditions in Eq.~(\ref{Eq:scalarbc}),
in the limit in which we remove the UV regulator (boundary) at $\r_{2}$.

We start the analysis from the DW solutions.
The result of the numerical study of the fluctuations for the $(\phi_-,{\cal A}_-)$ 
solutions  is displayed in Fig.~\ref{Fig:spectraDWn}, as a function of the parameter 
$\phi_o$ characterising this one-parameter family.
We find it convenient to normalise the masses $M$ of the spin-0 (blue disks) 
and spin-2 states (red circles) so that 
the lightest tensor mode has unit mass.

For any finite value of $\phi_o$ the spectrum is characterised by an unremarkable discrete 
sequence of states, and by the existence of a tachyon, which signals a fatal instability in the background solutions. 
Only in the strict limit $\phi_o\rightarrow -\infty$
does the tachyon become exactly massless. In the same limit, the spectrum degenerates to a  gapped continuum,
in all the channels, for $M^2>1$ (in units of the lightest tensor mode).
 This result reproduces the results  quoted from the 
 literature in Appendix~\ref{Sec:susy}, for the $n=2$ case, confirming 
 that the gauge-invariant formalism we adopt,  the choices of boundary 
 conditions we impose, and the numerical strategy we deploy combine to correctly identify all the 
 poles of the relevant 2-point correlation functions.
 
By comparing the gauge-invariant spin-0 fluctuations  to the probe approximation
 (the black diamonds in Fig.~\ref{Fig:spectraDWn}),
we clearly see that  the probe approximation
  fails most completely to capture the lightest (tachyonic) masses, for all values of $\phi_o$, so that we can establish
  that the dilaton component is always important in such spin-0 objects.
   Mixing effects become prevalent for negative $\phi_o$.
  For large and positive values of $\phi_o$, the probe approximation captures well the
   excited scalar states, and hence yields a clear, unambiguous identification of the dilaton with the tachyon.

Fig.~\ref{Fig:spectraDWp} displays the result of the study of the fluctuations for the DW solutions
$(\phi_+,{\cal A}_+)$. These include the supersymmetric $n=4$ case in Appendix~\ref{Sec:susy}  
(marked for convenience by a vertical dashed line in the figure.)
The special limiting case of $\phi_o\rightarrow +\infty$ is reached asymptotically at the right-hand side of the figure.
The symbols (and colors) follow the same conventions as in Fig.~\ref{Fig:spectraDWn}.

One  difference appears immediately evident: there is a region of parameter space
 ($\phi_o>-\frac{1}{2}\sqrt{\frac{3}{2}}\log\left(\frac{4}{3}\right)$),
bounded by the supersymmetric $n=4$ solution, 
over which all the scalar states have positive-definite mass squared.
(We also saw in Sect.~\ref{Sec:positiveDWSolutions}, and particularly in 
Sect.~\ref{Sec:specialpositiveDWSolutions}, that solutions of this type have a milder singularity.)

Once more the spectrum of the supersymmetric background is in  agreement with the literature,
which further confirms that our numerical strategy is reliable. 
The spectrum for  $\phi_o<-\frac{1}{2}\sqrt{\frac{3}{2}}\log\left(\frac{4}{3}\right)$ always contains a tachyon, 
followed by a light scalar state and a densely packed sequence of heavy excitations. In the limit
$\phi_o\rightarrow -\infty$, the spectrum agrees  with the case of the supersymmetric solution 
$(\phi_2,{\cal A}_2)$, except for the addition of a tachyon. Indeed, this superficially surprising feature
can be explained by close examination of the stream plot in Fig.~\ref{Fig:stream}, from which one can see that
it is possible to choose boundary conditions for the $(\phi_+,{\cal A}_+)$ solutions 
that yield a trajectory approaching the special (supersymmetric) limit $(\phi_2,{\cal A}_2)$ with $\phi_o\rightarrow -\infty$.
Yet,  eventually all the  $\phi_+(\r)$ solutions
turn positive (and divergent), close enough to the end of the space;  the tachyon
emerges as an unavoidable consequence of the intrinsic instability of these flows.

The comparison with the probe approximation is instructive: the tachyon is never captured by the 
probe approximation, which rather produces an arbitrary number of negative-mass-squared states, 
depending on $\phi_o$. Conversely, in the region of large and positive $\phi_o$ the probe approximation
 highlights that an infinite number of scalars mix with the dilaton.

The spectrum of confining solutions with $\phi=0$ has been computed
in Ref.~\cite{Brower:2000rp}, although  only after truncating the tower of excitations of $\phi$.
This background is sometimes called QCD$_3$ (with  abuse of language) 
in the literature.
In units of the lightest spin-2 excitation 
the spectrum of mass of the tensors 
($T_3$ in Table~4 of Ref.~\cite{Brower:2000rp})
is reported to be
\beqs
M_2&=&1\,,\,1.73\,,\,2.44\,,\,.3.15\,,\,3.86\,,\,4.56\,,\,\cdots\,,
\eeqs
and for the scalars ($S_3$ in Table~4 of Ref.~\cite{Brower:2000rp}):
\beqs
M_0&=&0.69\,,\,1.62\,,\,2.37\,,\,.3.10\,,\,3.81\,,\,4.53\,,\,\cdots\,.
\eeqs

We extend the numerical study to
 the whole one-parameter scion of solutions characterised by $\phi_o$, 
 and retain both fluctuations of $\phi$ and $\chi$.
The resulting spectrum 
is displayed in Fig.~\ref{Fig:spectraConf}. The masses of scalars (blue disks) and
tensors (red circles) are plotted as a function of  $\phi_I$.
We also display the result of applying the probe approximation to the treatment of the 
scalars (black diamonds).
For $\phi_I=0$, the tensor masses, as well as the masses of half the scalars 
(the second, fourth, six, eight, \dots) 
are in excellent agreement with Ref.~\cite{Brower:2000rp}, confirming
for the third time
 the robustness of our procedure.
Yet,  the truncation adopted in Ref.~\cite{Brower:2000rp} misses the lightest of the scalar states, 
which can be decoupled only for $\phi=0$---in this case, the probe approximation is accurate for the first, third, fifth, \dots, 
scalar states, but only within  a narrow 
range around $\phi=0$.

The  main feature that emerges is that the spectrum is positive definite
only for $\phi_I>\phi_I^{\ast}\simeq -0.52$. For large and negative values of $\phi_I$, we find the 
emergence of a tachyon, signaling the appearance of an instability. 
The probe approximation fails 
to capture the features of the spectrum, even at the qualitative level, yielding an unphysical proliferation of
tachyons.

In summary, in the case of the confining solutions,
and  for $\phi_I>\phi_I^{\ast}\simeq -0.52$, the solutions are regular (the curvature invariants
computed in $D=5$ and $D=10$ dimensions are all finite) and smooth 
(there is no conical singularity at the end of space),
the spectrum is positive definite, and the calculation of the Wilson loop via the dual
gravity prescription leads to the linear potential
expected in a confining field  theory (in three dimensions).
All of these properties are preserved all the way along the scion of confining solutions until
the critical value $\phi_I^{\ast}$, in proximity of which the lightest scalar separates from the rest of the spectrum,
and becomes arbitrarily light, before turning into a tachyon. This light state, as shown by the
 probe approximation, has a substantial overlap with the dilaton,
and couples to the trace of the energy-momentum tensor of the dual confining theory.

\section{Free energy and stability analysis}
\label{Sec:Comparison}

Because we regulate the theory by introducing two boundaries $\r_1$ and $\r_2$
 in the radial direction of the geometry,
the complete action in $D=5$ dimensions must include also boundary-localised terms: 
\beqs
{\cal S}={\cal S}_{5}+\sum_{i={1,2}}(-)^i\int\di^3x\di\eta
\sqrt{-\tilde{\hat{g}}}\left[\frac{K}{2}+\lambda_i\right]\Bigg|_{\r_i}.
\eeqs
The Gibbons-Hawking-York (GHY) term is
 proportional to the extrinsic curvature $K$, and  $\lambda_i$ are boundary potentials.
We choose $\lambda_1=-\left.\partial_{\rho} A\right|_{\r=\r_1}$,
and, as in Ref.~\cite{Bianchi:2001kw} (see also Eq.~(3.66) of Ref.~\cite{Papadimitriou:2004rz}),
\beqs
\lambda_2&=&-\frac{3}{2}-\phi^2\left(1+\frac{1}{\log(k^2 z_2^2)}\right)\,,
\eeqs
where $z_2\equiv e^{-\r_2}$, and the freedom in the choice of  $k$ 
reflects the scheme-dependence of the result. 

The explicit 
appearance of the term containing the unphysical constant $k$ in this result is a peculiarity of this model, 
distinguishing it from  those in Refs.~\cite{Elander:2020ial,Elander:2020fmv}.
It is due to the mass of the scalar field corresponding to
the deforming field theory operator exactly saturating the BF unitarity bound.
In this sense, this model is a more direct realisation of the ideas exposed in
Ref.~\cite{Kaplan:2009kr}, where the proximity to the BF bound is the starting point of the analysis.
As we shall see shortly, though, our results here 
are qualitatively similar to those in Refs.~\cite{Elander:2020ial,Elander:2020fmv}.

The need for counter-terms that are quadratic in $\phi$, and their scheme dependence, imply that the 
concavity theorems do not hold for the free energy of this system.
The free energy  $F$ and its density ${\cal F}$ are defined as
\beqs
F &\equiv&-\lim_{\r_1\rightarrow \r_o}\lim_{\r_2\rightarrow +\infty} {\cal S}^{\rm on-shell}\,\equiv\,
\int \di^3x\di\eta\, {\cal F}\,,
\eeqs
and by using the equations of motion,
supplemented by the observation that Eq.~(\ref{Eq:conserved}) defines a conserved quantity along 
 the radial direction $\rho$, we find
\beqs
{\cal F}&=&
-\lim_{\r_2\rightarrow +\infty} \left.\frac{}{}e^{3A-\chi}
\left(\partial_{\rho}A\frac{}{}+\lambda_2
\right)\right|_{\r_2}\,.
\eeqs
We can now use the UV expansions,  take the $e^{-\r_2}\rightarrow 0$ limit, and arrive at
 \beqs
 {\cal F}&=&
 \frac{1}{18}e^{3A_U-\chi_U}\left(\frac{}{}2\phi_2^2+9\phi_2\phi_{2l}
 -2\phi_{2l}^2\frac{}{}
 +\right.\\
 &&\left.\nonumber
 +24\chi_4
 -9\phi_{2l}^2 \log(k)\frac{}{}\right)\,.
 \label{Eq:free}
 \eeqs
For  the DW solutions, further simplifications yield 
 \beqs
 {\cal F}^{(DW)}&=&\frac{1}{8}e^{\frac{8}{3}A_U}\Big(4\phi_2-\phi_{2l}-4\phi_{2l}\log(k)\Big)\phi_{2l}\,.
 \eeqs
 
Along the   lines of  Ref.~\cite{Elander:2020ial},
we find it convenient to define a scale $\Lambda$ as follows~\cite{Csaki:2000cx}:
\begin{equation}
\label{Eq:Scale}
\Lambda^{-1}\equiv
\int_{\rho_o}^{\infty}\di\rho\,e^{ \chi(\rho)-A(\r)}\,.
\end{equation}
While this is not a unique choice, its simplicity and universality gives it a 
practical value for our applications.
In the calculation of the free energy density, we set $k=\Lambda$,
as this quantity scales  with dilatations in the same way as $z^{-1}$.

\begin{figure}[t]
	\centering
	\begin{picture}(200,290)
	\put(0,145){\includegraphics[width=.44\textwidth]{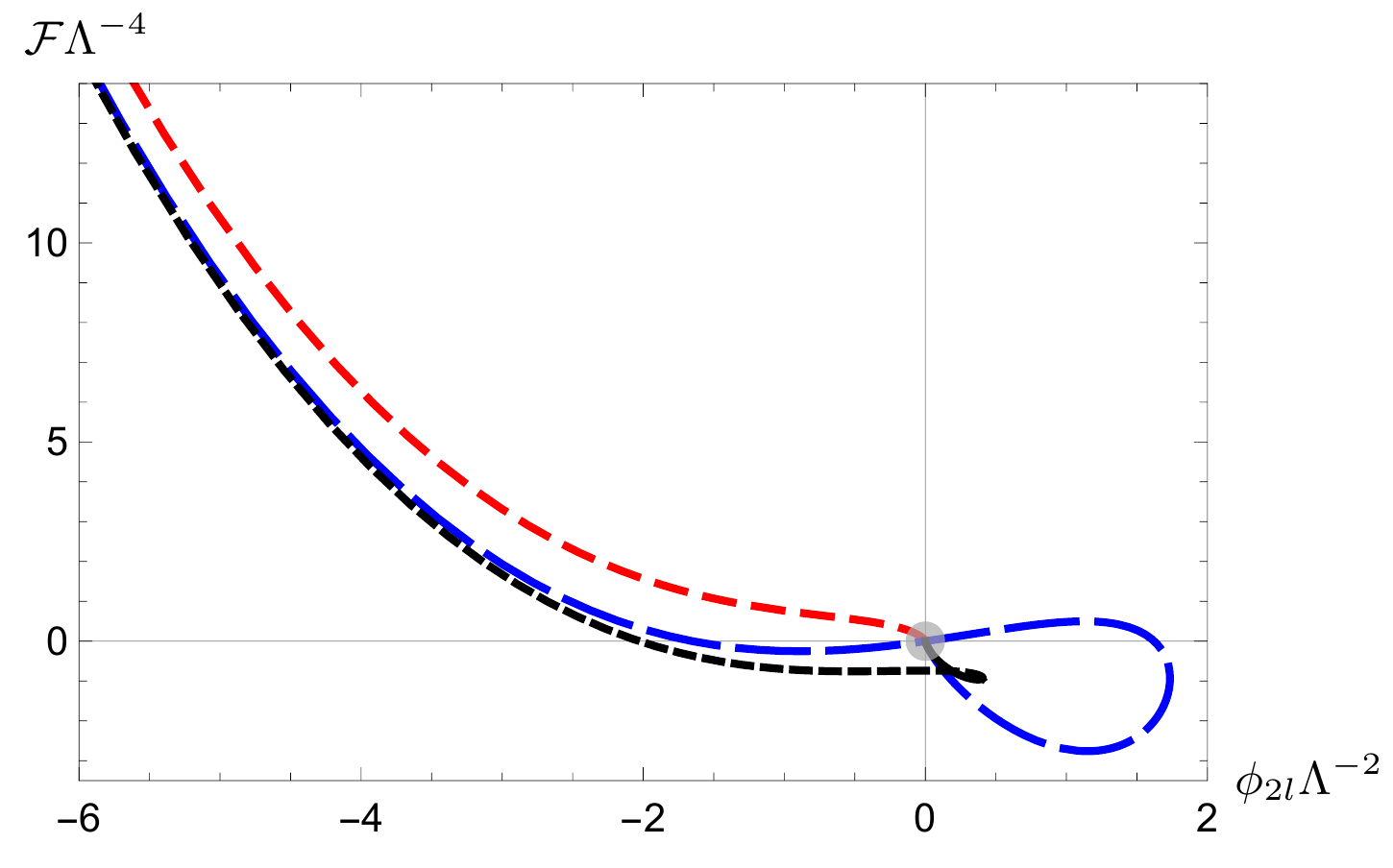}}
	\put(0,0){\includegraphics[width=.44\textwidth]{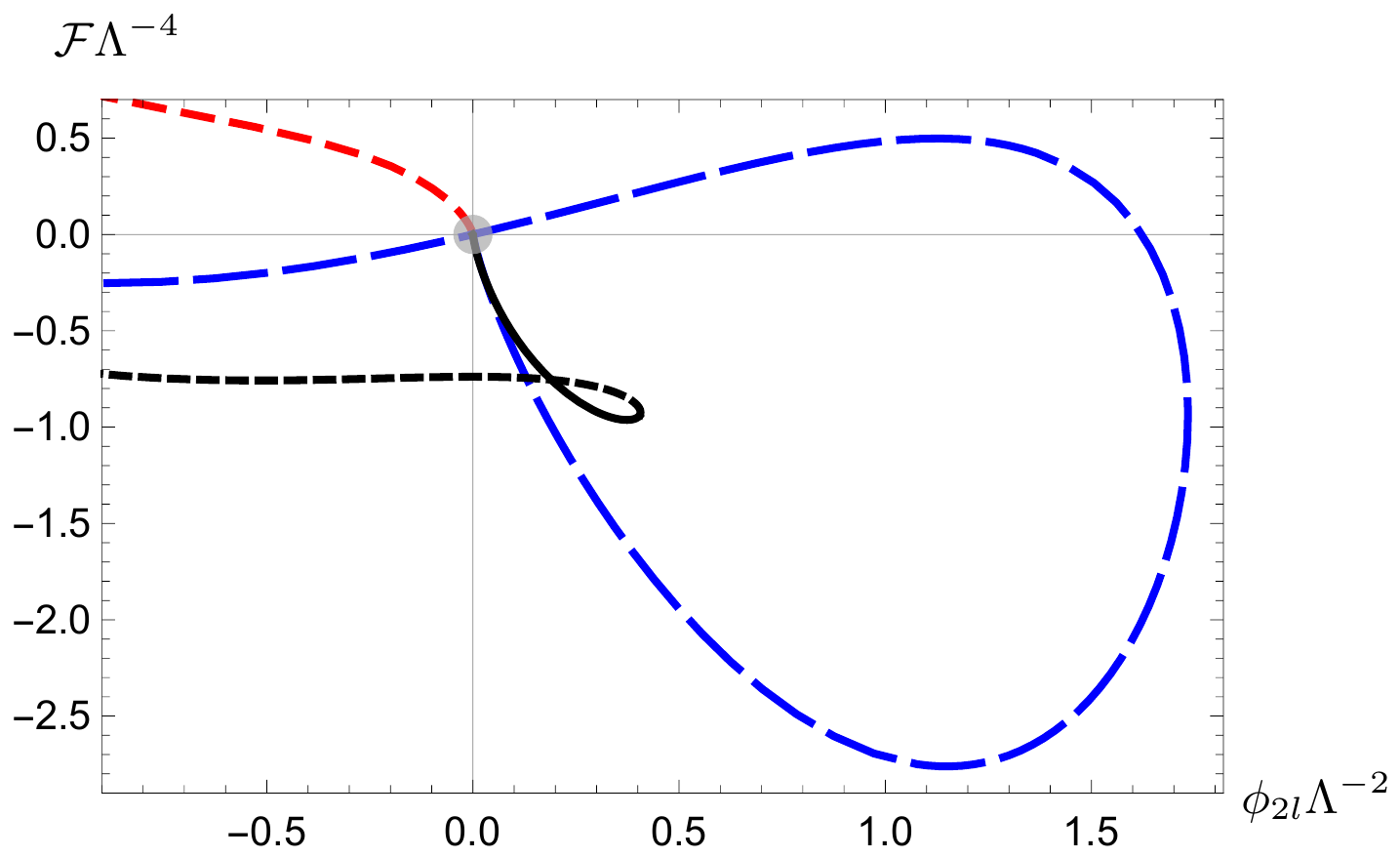}}
	\end{picture}
	\vspace{0mm}
	\caption{The free energy density ${\cal F}\Lambda^{-4}$, defined in Eq.~(\ref{Eq:free}),
	expressed in units of the 
	scale-setting parameter $\Lambda$ defined in Eq.~(\ref{Eq:Scale}), plotted as a function
	of the deformation parameter $\phi_{2l}\Lambda^{-2}$, for the various classes of solutions considered in the paper.
	For  the confining solutions, 
	the tachyonic backgrounds are denoted by continuous (black) lines,
	while backgrounds with  positive-definite mass spectrum are represented by the (black) short-dashed line.
	The long-dashed (blue) lines are the positive DW solutions,  the dashed (red) lines are the negative DW solutions,
	while the grey disk represents the supersymmetric solutions.
	The bottom panel is a detail of the top one.
		\label{Fig:Free}}
\end{figure}  

We display in Fig.~\ref{Fig:Free} the result of the calculation of ${\cal F}\Lambda^{-4}$ as a function of 
the source $\phi_{2l}\Lambda^{-2}$, for the three classes of negative DW, positive DW and confining solutions.
For negative values of $\phi_{2l}\Lambda^{-2}$, as we saw 
the regular confining solutions have positive definite spectrum (as $\phi_I>\phi_I^{\ast}$), and furthermore
their free energy  is the lowest among the solutions we considered.
When $\phi_{2l}\Lambda^{-2}$ is positive, but small, we still find regular confining solutions, 
but the lightest scalar state has lower mass, which vanishes when $\phi_I=\phi_I^{\ast}\simeq -0.52$,
after which it turns tachyonic. There is hence a regime of parameter space in which the lightest scalar 
has suppressed mass.
But these solutions are metastable: the positive DW solutions (despite being singular)
have lower free energy when $\phi_{2l}\Lambda^{-2}>\phi_{2l}^{\star}\Lambda^{-2} \simeq 0.13$,
the critical value identified by the crossing in Fig.~\ref{Fig:Free} (corresponding to $\phi_I^{\star}\simeq -0.067$)
Once more, as in the models in  Refs.~\cite{Elander:2020ial,Elander:2020fmv}, 
we find that the lightest scalar can be identified with a parametrically light
dilaton along the metastable solutions. In the stable solutions the lightest scalar
is still an approximate dilaton, but not parametrically light.

\section{Conclusion and outlook}
\label{Sec:Conclusions}

We studied new classes of background solutions  
of maximal supergravity in $D=5$ dimensions, truncated to retain only one
scalar field. This is the theory related to the dual of the Coulomb 
branch of ${\cal N}=4$ SYM.
We focused on solutions that are regular,  have positive definite spectrum, and can be interpreted as 
the gravity dual of confining field theories in three dimensions.
We found evidence that the lightest scalar state is an approximate dilaton, and can be made parametrically light,
in a region of parameter space in which these new regular solutions are metastable.

The study confirms, in a lower-dimensional simple setting, for a well known example of gauge-gravity duality
related to the study of ${\cal N}=4$ SYM,  the qualitative features that emerged in 
the models in  Refs.~\cite{Elander:2020ial,Elander:2020fmv}.
We notice the emergence of a first-order phase transition separating the metastable 
from the stable portions of the parameter space of the new confining solutions.
As in Refs.~\cite{Pomarol:2019aae,Gorbenko:2018ncu,Gorbenko:2018dtm},
the approximate dilaton is not parametrically light in the stable solutions,
confirming this generic feature also in  confining theories in three dimensions.

Further exploration of the catalogue of supergravity theories
will possibly help to understand whether the aforementioned results are universal or model dependent.
Of particular interest would be to ascertain
whether it is possible, and under what conditions, 
 to find systems for which the phase transition is weak enough to render
the dilaton parametrically light already in the stable region of parameter space,
in proximity of the phase transition itself. It would also be interesting to see whether systems exist for which the phase transition is of second order.

\begin{acknowledgments}
MP would like to thank Carlos Nunez for a  useful discussion.
The work of MP has been supported in part by the STFC Consolidated Grants  ST/P00055X/1
and ST/T000813/1. MP has also received funding from the European Research Council (ERC) under the
European Union Horizon 2020 research and innovation programme under grant agreement No 813942. 
JR is supported by STFC, through the studentship ST/R505158/1. 
\end{acknowledgments}

\appendix
\section{The two supersymmetric solutions}
\label{Sec:susy}

The first-order Eqs.~(\ref{Eq:FirstOrderA})  can be solved exactly by changing variable according to
 $\partial_{\rho}=e^{\frac{\phi}{\sqrt{6}}}\partial_{\tau}$.
The first-order equations are then
\beqs
\partial_{\tau}\phi(\tau)&=&-\frac{4}{\sqrt{6}}\sinh\left(\sqrt{\frac{3}{2}}\phi(\tau)\right),\\
\partial_{\tau}{\cal A}(\tau)&=&\cosh\left(\sqrt{\frac{3}{2}}\phi(\tau)\right)+\nonumber\\
&&-\frac{1}{3}\sinh\left(\sqrt{\frac{3}{2}}\phi(\tau)\right).
\eeqs
There are two exact solutions~\cite{Brandhuber:1999hb,Brandhuber:1999jr,Freedman:1999gk}.
The case of $n=2$ is given by
\beqs
\phi_2(\tau)&=&-\frac{4}{\sqrt{6}}{\rm arctanh}\left(e^{-2(\tau-\tau_o)}\right)\,,
\label{Eq:phisusy1}\\
{\cal A}_2(\tau)&=&{\cal A}_o+\tau-\tau_o-\frac{1}{3}{\rm arctanh}\left(e^{-2(\tau-\tau_o)}\right)
+\nonumber\\
&&
+\frac{1}{2}\log\left(1-\frac{}{}e^{-4(\tau-\tau_o)}\right)\,,
\label{Eq:aAsusy1}
\eeqs
and the $n=4$ case by
\beqs
\phi_4(\tau)&=&\frac{4}{\sqrt{6}}{\rm arctanh}\left(e^{-2(\tau-\tau_o)}\right)\,,
\label{Eq:phisusy2}\\
{\cal A}_4(\tau)&=&{\cal A}_o+\tau-\tau_o
+\frac{1}{3}{\rm arctanh}\left(e^{-2(\tau-\tau_o)}\right)
+\nonumber\\
&&
+\frac{1}{2}\log\left(1-\frac{}{}e^{-4(\tau-\tau_o)}\right)\,,
\label{Eq:aAsusy2}
\eeqs
where $\tau_o$ and ${\cal A}_o$ are two integration constants.

The two classes of supersymmetric solutions can be rewritten
as expansions valid for $0<\r-\r_o \ll 1$:
\beqs
\phi_2(\r)&=&
-\sqrt{\frac{3}{8}}\log\left(\frac{9}{4}\right)
+\sqrt{\frac{3}{2}}\log(\r-\r_o)+\label{Eq:phis2}\\
&&
\nonumber
-\sqrt{\frac{2}{243}}(\r-\r_o)^3
+\frac{17}{1701}\sqrt{\frac{2}{3}}(\r-\r_o)^6\,+\cdots\,,\\
{\cal A}_2(\r)&=&
\label{Eq:aAs2}
{\cal A}_I+\log(\r-\r_o)+\frac{2}{27}(\r-\r_o)^3+
\nonumber\\
&&
-\frac{22}{5103}(\r-\r_o)^6\,+\cdots\,.
\eeqs
and 
\beqs
\label{Eq:phis2+}
\phi_{4}(\r)&=&-\frac{1}{2}\sqrt{\frac{3}{2}}\log\left(\frac{4}{3}\right) -\frac{1}{2} \sqrt{\frac{3}{2}}\log(\r-\r_o)+
\\ \nonumber &&
+\frac{4}{15}\sqrt{2} (\r-\r_o)^{3/2}-\frac{28}{225}\sqrt{\frac{2}{3}}(\r-\r_o)^3\,+\cdots\,,\\
\label{Eq:aAs2+}
{\cal A}_{4} (\r) &=&{\cal A}_I+\frac{1}{4}\log(\r-\r_o)
+\frac{16}{15\sqrt{3}}(\r-\r_o)^{3/2}+
\nonumber\\
&&
-\frac{52}{675} (\r-\r_o)^3 +\cdots \,.
\eeqs
The $n=2$ case is the limit $\phi_o\rightarrow -\infty$ of the general DW solutions
in Eqs.~(\ref{Eq:phiDWn}) and~(\ref{Eq:aADWn}), while the $n=4$ case  
is recovered with the choice $\phi_o=-\frac{1}{2}\sqrt{\frac{3}{2}}\log\left(\frac{4}{3}\right)$
in Eqs.~(\ref{Eq:phiDWp}) and~(\ref{Eq:aADWp}).

\begin{figure}[b]
	\centering
	\begin{picture}(200,270)
\put(0,135){\includegraphics[width=.38\textwidth]{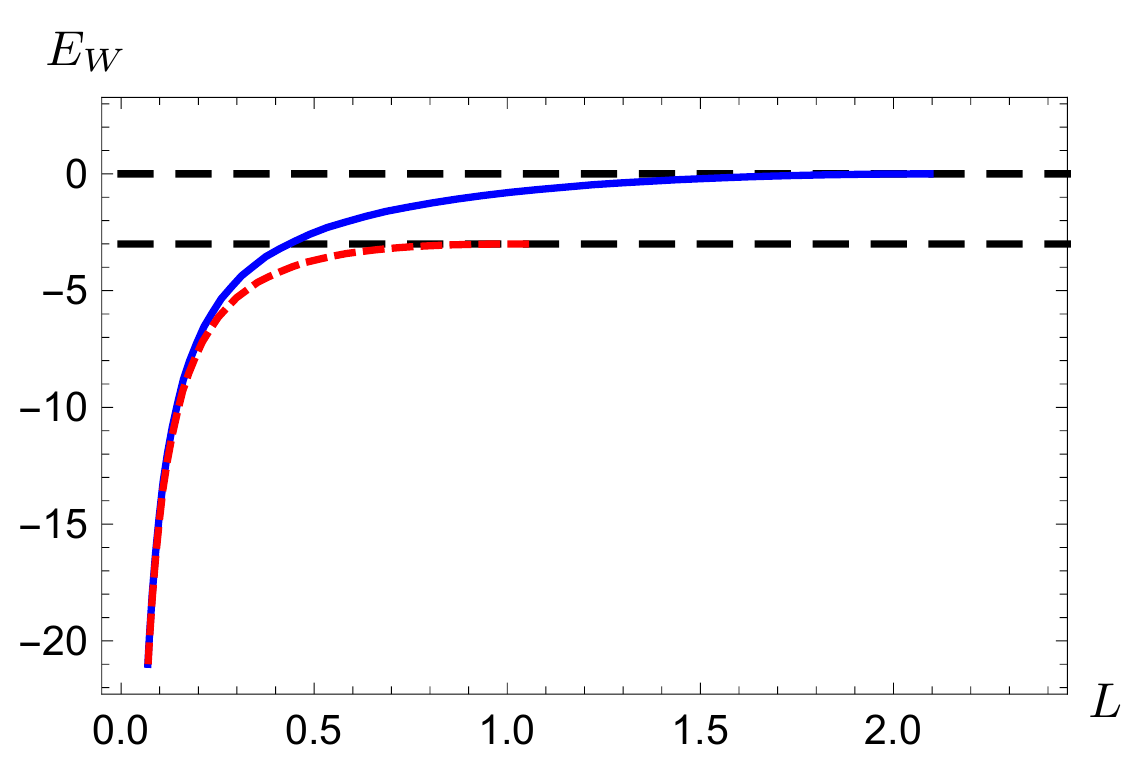}}
\put(6,0){\includegraphics[width=.38\textwidth]{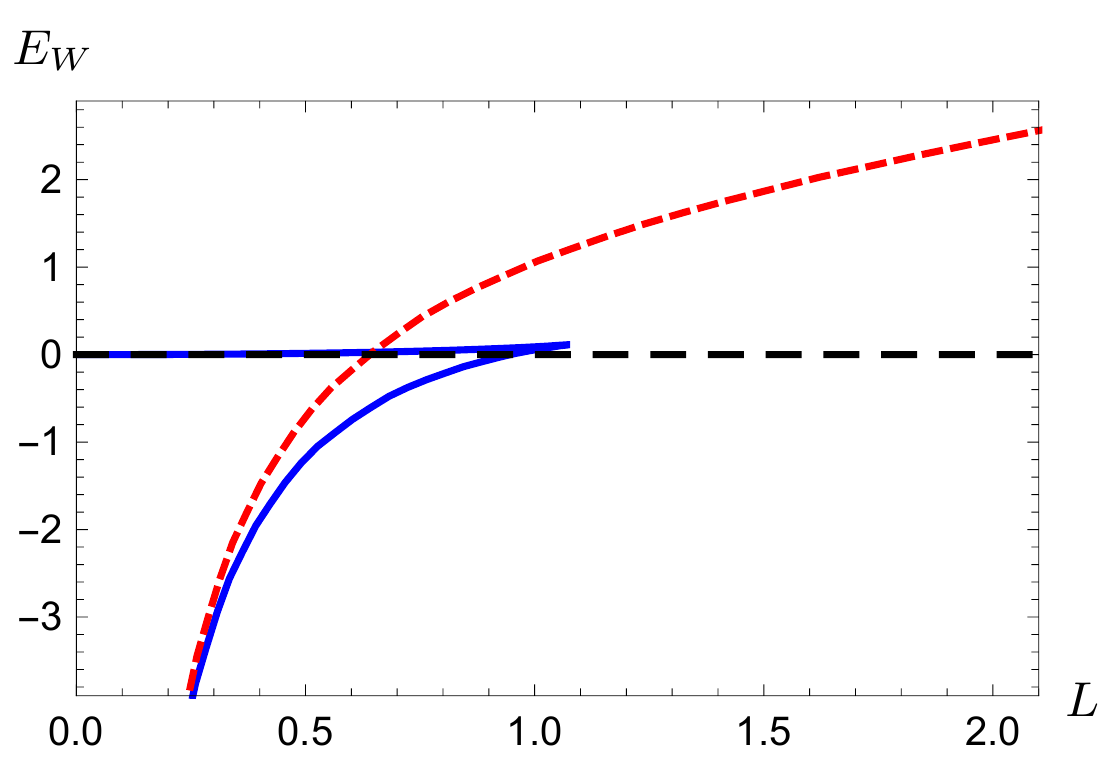}}
	\end{picture}
	\vspace{0mm}
	\caption{Energy $E_W$ as a function of the separation $L$, for supersymmetric solutions 
	 $n=2$, $(\phi_2,{\cal A}_2)$ (top) and $n=4$,  $(\phi_4,{\cal A}_4)$ (bottom),
	 and chosen to have ${\cal A}_I=0$.
	Continuous (blue) lines correspond to configurations with $\theta=0$, short-dashed (red) lines depict 
	configurations with $\theta=\pi/2$, and long-dashed (black) lines represent configurations reaching 
	exactly the end of space.
	\label{Fig:Wilson}}
\end{figure}  

Both solutions exhibit a naked  singularity  in $D=5$ dimensions,
 which softens in $D=10$ dimensions. For
 the $n=2$ solutions  $(\phi_2,{\cal A}_2)$ we find
\beqs
{\cal R}_{10,\hat{M}\hat{N}}{\cal R}_{10}^{\,\,\,\,\,\hat{M}\hat{N}}&=&
\frac{135 }{4\cos^2(\theta)(\r-\r_o)^{3}}
\label{Eq:RicciEquator2}
+\cdots\,.
\eeqs
For the ($n=4$) solutions given by $(\phi_4,{\cal A}_4)$ the behaviour of this invariant is milder:
\beqs
{\cal R}_{10,\hat{M}\hat{N}}{\cal R}_{10}^{\,\,\,\,\,\hat{M}\hat{N}}&=&
\frac{5\sin^4(2\theta)}{8\sin^{10}(\theta)}
\label{Eq:RicciEquator4}
+\cdots\,.
\eeqs
The singularity at
the equator of $S^5$ displayed by the curvature invariant in 
Eqs.~(\ref{Eq:RicciEquator2}) and~(\ref{Eq:RicciEquator4}) 
 at the end of space signals  the incompleteness of 
the supergravity description in both  supersymmetric cases.

The Wilson loops for the supersymmetric solutions have been computed in Ref.~\cite{Brandhuber:1999jr}.
We display our result in Fig.~\ref{Fig:Wilson}, as a test of our procedure.
In the case of the $n=2$ solutions $(\phi_2,{\cal A}_2)$,
for both $\theta=0,\pi/2$  we find a monotonic potential, and a maximum value of $L=L_{\rm max}$.
In the case $n=4$ of $(\phi_4,{\cal A}_4)$, there is a very major difference 
between the two cases with $\theta=0,\pi/2$. The case of $\theta=0$
 displays the features expected by a first order phase transition,
with long-distance screening. 
Conversely, for $\theta=\pi/2$
we find a linear potential at asymptotically large $L$, 
which is unbounded;
this   configuration has higher energy than the $\theta=0$ one.

For the $n=2$ solutions  $(\phi_2,{\cal A}_2)$ in Eqs.~(\ref{Eq:phisusy1}) 
and~(\ref{Eq:aAsusy1}), 
the 2-point function of the operator $\co$ dual to the scalar $\phi$ can be found in 
Section~3.3 of Ref.~\cite{Papadimitriou:2004rz} (see also Eq.~(8.6) of Ref.~\cite{Bianchi:2001kw}):
\beqs
\langle {\cal O}(q){\cal O}(-q)\rangle=
\frac{16}{3q^2}-4\left(\psi(a(q)+1)\frac{}{}-\psi(1)\right),
\eeqs
and for the tensors
\beqs
\langle T_{\mu\nu}T_{\rho\sigma}\rangle
\propto\frac{q^2}{2\kappa^{\prime}}
\Big[\frac{1}{3}-\frac{q^2}{2}\Big(\psi(a(q)+1)\frac{}{}-\psi(1)\Big)\Big],
\eeqs
where $\psi$ is the digamma function, $q$ the four-momentum
in Euclidean signature, $\kappa^{\prime}$ is a constant,
and $a$ is
\beqs
a(q)&\equiv&-\frac{1}{2}+\frac{1}{2}\sqrt{1+{q^2}{}}\,.
\label{Eq:a}
\eeqs
The scalar correlator displays a massless pole, a gap and a continuum cut,
the tensor differs by the absence of the massless state.

The $n=4$ solutions $(\phi_4,{\cal A}_4)$ in Eqs.~(\ref{Eq:phisusy2})
and~(\ref{Eq:aAsusy2}), have a discrete spectrum, for example described in Eq.~(26) of Ref.~\cite{Freedman:1999gk}.
With $j=0, 1, \cdots$, the spectrum is given by 
$M\propto \sqrt{j(j+1)/2}\simeq 0, 1, 1.7, 2.5, 3.2, 
\cdots$. 
The spectrum of tensors can be found for example in Eq.~(45) of Ref.~\cite{Brandhuber:1999hb},
according to which it agrees with that of the scalars `\dots to an extremely good approximation \dots',
 but for the fact that there is no zero mode.

The results of studying the Wilson loops agree with Figs.~1, 4, and~5 of Ref.~\cite{Brandhuber:1999jr}.
We  relied on numerical solutions guided by the asymptotic IR expansions, 
rather than using  the exact solutions as in Ref.~\cite{Brandhuber:1999jr}.
Our numerical  study of the spectrum
 yields numerical results in splendid agreement with pre-existing
 calculations, as can be seen in Figs.~\ref{Fig:spectraDWn} and~\ref{Fig:spectraDWp}.
These results and their agreement with earlier studies of the supersymmetric solutions
confirm the robustness of our formalism and 
numerical strategy.

\section{Expansions for the fluctuations}
\label{Sec:exp}

In the numerical calculation of the spectra, we used the asymptotic expansions of the 
gauge-invariant fluctuations, as a way to optimise the decoupling of spurious 
cutoff effects present at finite $\r_2$.
In the case of DW solutions, we find that we can expand the physical fluctuations as follows:
\beqs
\mathfrak{a}^\phi &=& \mathfrak{a}^\phi_{2l} \log(z) z^2 + \mathfrak{a}^\phi_2 z^2 + {\cal O}(z^4)\,,\\
\mathfrak{e}^\mu_{\,\,\,\,\nu} &=& (\mathfrak{e}_0)^\mu_{\,\,\,\,\nu} \left(1 - \frac{e^{-8A_U} }{4} q^2 z^2 - 
\frac{ e^{-16A_U} }{16} q^4 \log(z)  z^4\right) + \nonumber\\
&& + (\mathfrak{e}_4)^\mu_{\,\,\,\,\nu} z^4 + {\cal O}(z^6)\,,
\eeqs
In these expressions, $\mathfrak{a}^\phi_{2l}$, $\mathfrak{a}^\phi_{2}$, 
$(\mathfrak{e}_0)^\mu_{\,\,\,\,\nu}$, and $(\mathfrak{e}_4)^\mu_{\,\,\,\,\nu}$ are the 
integration constants governing the solutions of the second-order linearised equations,
half of which are determined by the boundary conditions.
In the probe approximation, the expansion for the scalars $\mathfrak{p}^{\phi}$ 
is of the same form, up to $ {\cal O}(z^4)$:
\beqs
\mathfrak{p}^{\phi} =  \mathfrak{p}^\phi_{2l} \log(z) z^2 + \mathfrak{p}^\phi_2 z^2  + {\cal O}(z^4)\,.
\eeqs

The confining solutions do not satisfy the DW conditions.
For the scalars we find
\beqs
\mathfrak{a}^\phi &=& \mathfrak{a}^\phi_{2l} \log(z) z^2 + \mathfrak{a}^\phi_2 z^2 + {\cal O}(z^4)\,,\\
\mathfrak{a}^\chi &=& \mathfrak{a}_0^\chi \left(1 - \frac{e^{-2A_U + 2 \chi_U}}{4} q^2 z^2 \frac{}{}
+\right.\\
&& \left.\nonumber
-\frac{e^{-4A_U +4\chi_U} }{16} 
q^4 \log(z)z^4\right)
+ \mathfrak{a}_4^\chi z^4 + {\cal O}(z^6)\,,
\eeqs
where $\mathfrak{a}^\phi_{2l}$, $\mathfrak{a}^\phi_{2}$, $\mathfrak{a}^\chi_{0}$, and $\mathfrak{a}^\chi_{4}$
are the free parameters.
For the probe approximation of the scalars, the free parameters are 
$\mathfrak{p}^\phi_{2l}$, $\mathfrak{p}^\phi_{2}$, $\mathfrak{p}^\chi_{0}$, and $\mathfrak{p}^\chi_{4}$, and
as a result of mixing in the second derivative of the potential in Eq.~(\ref{Eq:probe})
we have:
\beqs
\mathfrak{p}^\phi = \mathfrak{p}_{2l}^\phi \log(z) z^2 +\mathfrak{p}_2^\phi z^2 + {\cal O}(z^4)\,,\\
{ \mathfrak{p}^\chi = \mathfrak{p}_{2l}^\chi \log(z) z^2 +\mathfrak{p}_2^\chi z^2 + {\cal O}(z^4)}\,.
\eeqs
For the tensor fluctuations we find
\beqs
\mathfrak{e}^\mu_{\,\,\,\,\nu} &=& (\mathfrak{e}_0)^\mu_{\,\,\,\,\nu} \left(1 - \frac{1}{4} e^{-2A_U+2\chi_U} q^2 z^2 +
\right.\\
&&\left.
- \frac{1}{16} e^{-4A_U+4\chi_U} q^4  \log(z)  z^4\right) 
+ (\mathfrak{e}_4)^\mu_{\,\,\,\,\nu} z^4 + {\cal O}(z^6)\,.\nonumber
\eeqs


\end{document}